\documentclass[5p,twocolumn]{elsarticle}

\usepackage{hyperref}

\usepackage{threeparttable} 
\usepackage{graphicx}
\usepackage{epsfig}
\usepackage{amsmath}
\usepackage{amssymb}
\usepackage{natbib}

\journal{Journal of High Energy Astrophysics}

\newcommand{\chan}{\textit{Chandra}}
\newcommand{\swift}{\textit{Swift}}
\newcommand{\rxte}{\textit{RXTE}}
\newcommand{\xmm}{\textit{XMM-Newton}}
\newcommand{\inte}{\textit{Integral}}
\newcommand{\maxi}{\textit{MAXI}}

\newcommand{\nustar}{\textit{NuSTAR}}
\newcommand{\suzaku}{\textit{Suzaku}}

\newcommand{\Msun}{\mathrm{M}_{\odot}}
\newcommand{\lum}{\mathrm{erg~s}^{-1}}
\newcommand{\flux}{\mathrm{erg~cm}^{-2}~\mathrm{s}^{-1}}
\newcommand{\fluence}{\mathrm{erg~cm}^{-2}}
\newcommand{\cnts}{\mathrm{c~s}^{-1}}
\newcommand{\mdot}{\mathrm{M_{\odot}~yr}^{-1}}

\newcommand{\nh}{\mathrm{cm}^{-2}}
\newcommand{\distxmm}{(D/\mathrm{6.5~kpc})^{2}}
\newcommand{\distgrs}{(D/\mathrm{6.7~kpc})^{2}}

\newcommand{\xmmbron}{XMM J174457--2850.3}
\newcommand{\psr}{PSR J1023+0038}
\newcommand{\xss}{XSS J12270--4859}
\newcommand{\psrm}{PSR J1824--24521}

\newcommand{\saxrudy}{SAX J1750.8--2900}

\newcommand{\xte}{XTE J1701--462}
\newcommand{\maxisource}{MAXI~J0556--332}

\newcommand{\sgra}{Sgr~A$^{*}$}
\newcommand{\new}{Swift J174535.5--285921}

\newcommand{\ascabron}{AX~J1745.6--2901}

\newcommand{\grsbron}{GRS~1741--2853}
\newcommand{\ksbron}{KS~1741--293}
\newcommand{\magnetar}{SGR J1745--29}

\newcommand{\brontwee}{CXOGC~J174535.5--290124}
\newcommand{\brondrie}{CXOGC~J174540.0--290005}
\newcommand{\bronvier}{Swift~J174553.7--290347}

\newcommand{\bronvijf}{Swift~J174622.1--290634}

\newcommand{\bronacht}{CXOGC~J174538.0--290022}
\newcommand{\adcbron}{CXOGC~J174540.0--290031}

\newcommand{\bronnegen}{Swift J174535.5--285921}
\newcommand{\bronnegencxo}{CXOGC J174535.6--285928}
\newcommand{\othercxobron}{CXOGC~J174541.0--290014}

\def \mnras {MNRAS}
\def \apj {ApJ}
\def \apjs {ApJS}
\def \apjl {ApJL}
\def \aap {A\&A}
\def \nat {Nature}
\def \araa {ARAA}
\def \atel {ATel}

\def \pasj {PASJ}

\def \ssr {SSRv}
\def \physrep {Phys Rep}

\begin{document}

\begin{frontmatter}

\title{The Swift X-ray monitoring campaign of the center of the Milky Way}

\author[nde]{N. Degenaar\corref{cor1}}
\ead{degenaar@ast.cam.ac.uk}

\author[ams]{R. Wijnands}
\author[uom]{J.M. Miller}
\author[uom]{M.T. Reynolds}
\author[psu]{J. Kennea}
\author[gsfc]{N. Gehrels}

\cortext[cor1]{Corresponding author}

\address[nde]{Institute of Astronomy, University of Cambridge, Madingley Road, Cambridge CB3 OHA, UK}
\address[ams]{Anton Pannekoek Institute of Astronomy, University of Amsterdam, Science Park 904,1098 XH Amsterdam, The Netherlands}
\address[uom]{Department of Astronomy, University of Michigan, 1085 South University Avenue, Ann Arbor, MI  48109, USA}
\address[psu]{Department of Astronomy and Astrophysics, 525 Davey Lab, Pennsylvania State University, University Park, PA 16802, USA}
\address[gsfc]{Astrophysics Science Division, NASA Goddard Space Flight Center, Greenbelt, MD, USA}


\begin{abstract}
In 2006 February, shortly after its launch, \swift\ began monitoring the center of the Milky Way with the on board X-Ray Telescope using short 1-ks exposures performed every 1--4 days. Between 2006 and 2014 over 1200 observations have been obtained, accumulating to $\simeq$1.3~Ms of exposure time. This has yielded a wealth of information about the long-term X-ray behavior of the supermassive black hole \sgra, and numerous transient X-ray binaries that are located within the $25' \times 25'$ region covered by the campaign. In this review we highlight the discoveries made during these first nine years, which includes 1) the detection of seven bright X-ray flares from \sgra, 2) the discovery of the magnetar \magnetar, 3) the first systematic analysis of the outburst light curves and energetics of the peculiar class of very-faint X-ray binaries, 4) the discovery of three new transient X-ray sources, 5) exposing low-level accretion in otherwise bright X-ray binaries, and 6) the identification of a candidate X-ray binary/millisecond radio pulsar transitional object. We also reflect on future science to be done by continuing this \swift's legacy campaign, such as high-cadence monitoring to study how the interaction between the gaseous object `G2' and \sgra\ plays out in the future.
\end{abstract}


\end{frontmatter}

\section{Description of the program}
\begin{figure}[h!]
 \begin{center}
\includegraphics[width=8.7cm]{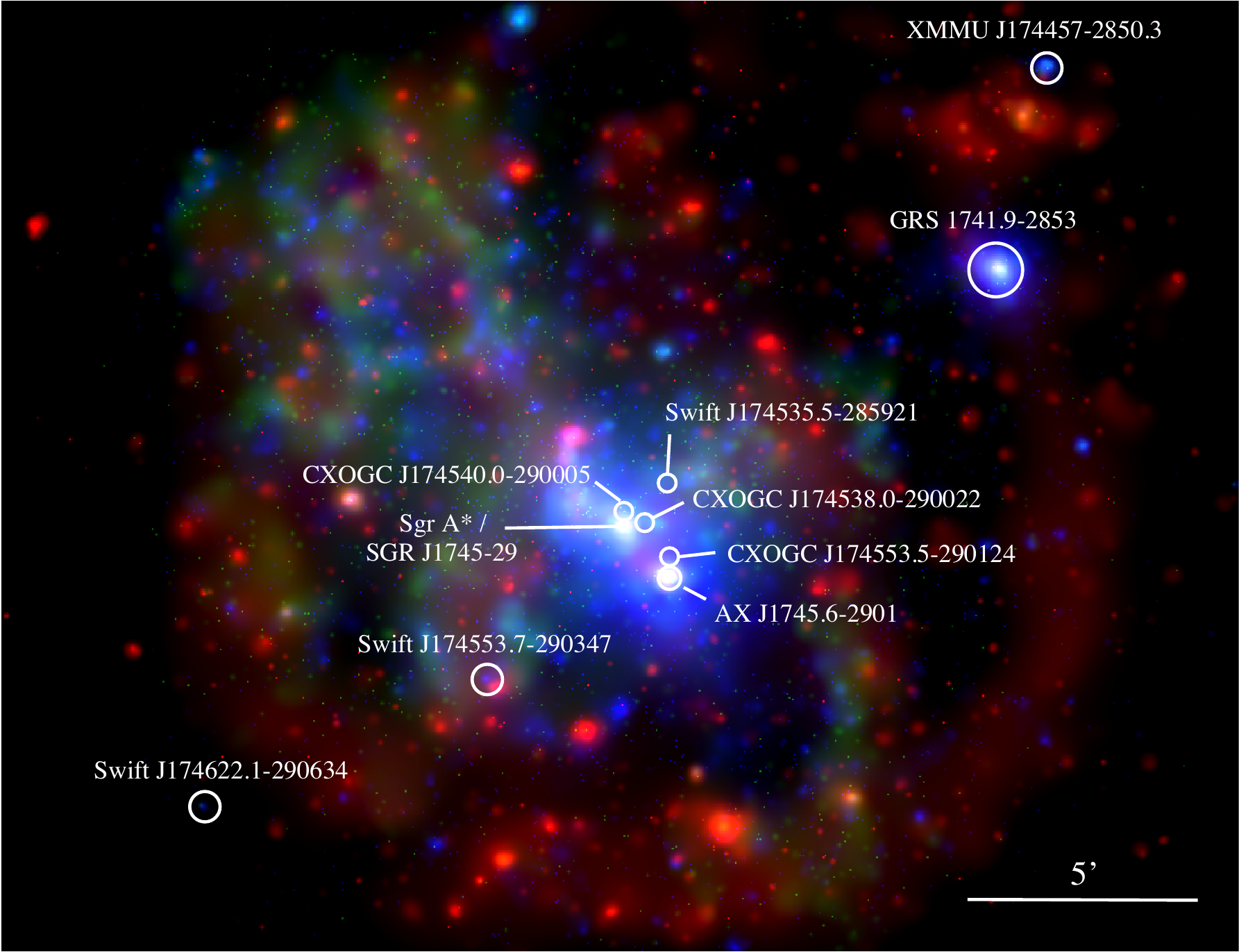}\vspace{-0.2cm}
    \end{center}
\caption[]{{Accumulated image of the XRT-PC data of the Galactic center using 1.3 Ms of data collected in 2006--2014 ($\simeq25'$ square). This representative-color image was constructed from 0.3--1.5 keV (red), 1.5--3.0 keV (green), and 3.0--10 keV (blue) images that were smoothed using the \textsc{ciao} tool \textsc{csmooth} and then combined with \textsc{ds9}. The markers indicate the location of the supermassive black hole \sgra\ and the magnetar \magnetar, as well as the 9 transient X-ray binaries that were seen active during the campaign.}}
\vspace{-0.4cm}
 \label{fig:image}
\end{figure}

Little over a year after the \swift\ satellite \citep[][]{gehrels2004} was launched in 2004 November, it embarked on a program to monitor the inner $\simeq25' \times 25'$ of the Milky Way using the on board X-Ray Telescope (XRT; \citep[][]{burrows05}). Starting in 2006 February, short $\simeq$1 ks X-ray snapshots of the Galactic center have been taken once every 1--4 days (see Table~\ref{tab:obs}). \swift\ can observe this region for $\simeq250$~days per year; the field is too close to the Moon for $\simeq$2--3 days per month and in November/December/January it is too proximate to the Sun. In 2006--2014 over 1200 X-ray observations were obtained in photon counting (PC) mode, amounting to $\simeq$1.3~Ms of exposure time (Table~\ref{tab:obs}). Figure~\ref{fig:image} displays an accumulated three-color XRT image of the Galactic center monitoring program.

\begin{table*}
\caption{Overview of \swift/XRT monitoring observations of the Galactic center.}
\begin{threeparttable}
\begin{tabular*}{0.99\textwidth}{@{\extracolsep{\fill}}ccccccc}
\hline
Year & Start Date & End Date & Total Observations & Total Exposure & Cadence  & Average Exposure \\ 
 &  &  &  & (ks) & (obs~day$^{-1}$)  & (ks~obs$^{-1}$) \\ 
\hline
1 & 2006 Feb 24 & 2006 Nov 1 & 197 & 262 & 0.8 & 1.3 \\	
2 & 2007 Feb 17 & 2007 Nov 1 & 173 & 172 & 0.8 & 1.0\\ 
3 & 2008 Feb 19 & 2008 Oct 30 & 162 & 200 & 0.6 & 1.2 \\ 
4 & 2009 May 17 & 2009 Nov 1 & 39 & 40 & 0.2 & 1.0 \\ 
5 & 2010 Apr 7 & 2010 Oct 31 & 64 & 71 & 0.3 & 1.1 \\ 
6 & 2011 Feb 4 & 2011 Oct 25 & 80 & 76 & 0.3 & 1.0 \\ 
7 & 2012 Feb 5 & 2012 Oct 31 & 80 & 74 & 0.3 & 0.9 \\ 
8 & 2013 Feb 3 & 2013 Oct 31 & 190 & 174 & 0.7 & 0.9 \\  
9 & 2014 Feb 2 & 2014 Nov 2 & 240 & 234 & 0.9 & 1.0 \\ 
\hline
1--9 &  2006 Feb 24 & 2014 Nov 2 & { 1226} & {1304} & { 0.5} & { 1.0} \\ 
\hline
\end{tabular*}
\label{tab:obs}
\begin{tablenotes}
\item[] Note. -- This overview concerns observations obtained in PC mode (updated from \citep[][]{degenaar2013_sgra}). In 2007 the daily monitoring observations were interrupted for 46 days between August 11 and September 26 \citep[][]{gehrels2007a,gehrels2007b}. 
\vspace{-0.3cm}
\end{tablenotes}
\end{threeparttable}
\end{table*}

The central part of the Milky Way has always been a prime target for X-ray missions as it is a very rich environment to study accretion onto compact objects. Not only does it harbor the central supermassive black hole Sagittarius A* (\sgra), there is also a large concentration of X-ray point sources toward the inner part of the Galaxy (e.g., \citep[][]{muno2009}). Many of these are X-ray binaries; binary star systems in which a stellar-mass black hole or a neutron star accretes matter from its companion. These are excellent laboratories to further our understanding of the physics of accretion, the properties of black holes and neutron stars, and stellar/binary evolution. 

Over the past decades, many different observing campaigns and many different satellites have targeted the Galactic center (e.g., \citep[][]{watson1981,pavlinsky1994,sidoli99,sidoli01,swank2001,sakano02,zand04,wijnands06,kuulkers07,muno2009,degenaar2012_gc,degenaar2013_sgra}). However, the Swift monitoring program is unique in many ways, and important for a variety of scientific goals. Firstly, the XRT provides better spatial resolution (arcseconds) and X-ray sensitivity (down to a 2--10 keV luminosity of $L_{X}$$\simeq$$10^{34}~\lum$ in a single 1-ks pointing) than wide-field monitors (e.g., \inte, \maxi, \rxte, \swift/BAT; arcminute resolution and a sensitivity of $L_{X}$$\simeq$$10^{36}~\lum$). Secondly, \swift\ is very flexible (compared to e.g., \chan, \xmm, \suzaku), allowing for many repeated observations and thereby providing unprecedented dense time coverage. 

During its nine year runtime, \swift's monitoring campaign of the Galactic center has led to many remarkable discoveries, which are reviewed in this article. For instance, the program allowed the detection of seven bright X-ray flares from \sgra\ (Section~\ref{subsec:flares}) and the discovery of the nearby (separated by only $\simeq$$2.4''$) magnetar \magnetar\ (Section~\ref{subsec:magnetar}). Furthermore, a total of 24 distinct accretion outbursts were detected from nine different transient X-ray binaries, of which three were newly discovered sources (Section~\ref{subsec:new}). The majority of these outbursts had peak luminosities of $L_{X}$$\lesssim$$10^{36}~\lum$ and trace a relatively poorly understood regime of accretion. The \swift\ program  has allowed for the first detailed and systematic analysis of such low-level accretion events (Sections~\ref{subsec:vf} and~\ref{subsec:low}). Moreover, the remarkable X-ray variability of one of these X-ray binaries suggests that it could possibly be a member of the recently emerged class of ``transitional'' neutron stars that switch between millisecond radio pulsar and X-ray binary manifestations (Section~\ref{subsec:trans}). In the next sections we describe these discovery highlights of \swift's Galactic center monitoring campaign in more detail, and we conclude with future prospects in Section~\ref{subsec:conclude}.\footnote{Throughout this work we quote uncertainties as 90\% confidence levels and assume source distances of 8~kpc unless noted otherwise.}

\begin{figure*}
 \begin{center}
\includegraphics[width=14.5cm]{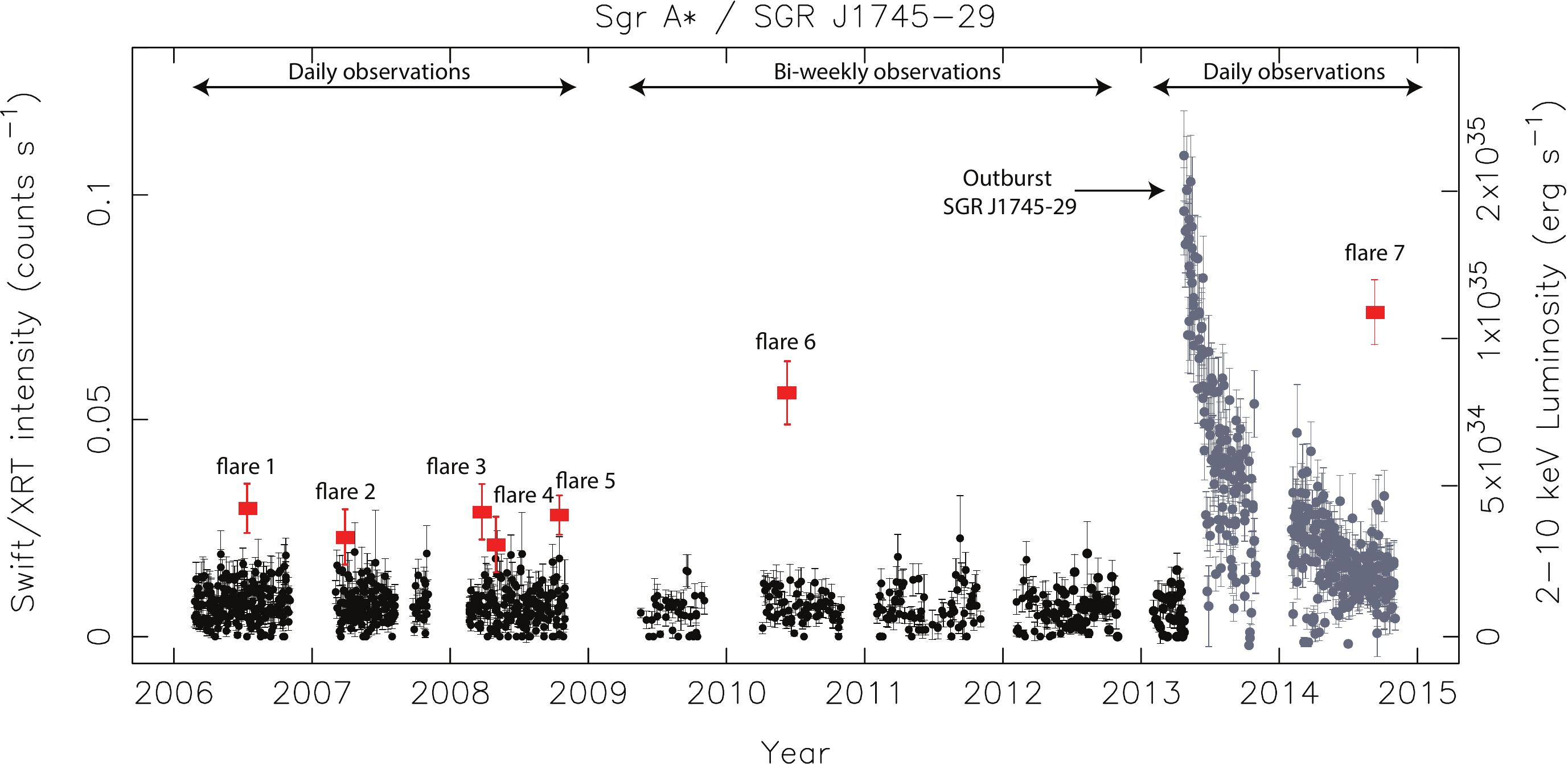} \vspace{-0.5cm}
    \end{center}
\caption[]{{Long-term XRT-PC count rate light curve of \sgra/\magnetar. The red squares indicate the seven X-ray flares detected from \sgra\ and the discovery outburst of \magnetar\ is indicated by the grey data points. The typical sampling rate along the campaign is indicated on top. Data gaps at the beginning/end of a year correspond to the Sun-constrained window and the interruption in 2007 corresponds to a short episode during which no observations could be obtained \citep[][]{gehrels2007a,gehrels2007b}. Note that this graph is binned per observation, whereas X-ray flares 1--5 were detected only during smaller data segments \citep[][]{degenaar2013_sgra}. Hence their true peak intensity was higher than reflected here.}}
 \vspace{-0.2cm}
 \label{fig:sgralc}
\end{figure*}

\section{Discovery highlights}

\subsection{Seven bright X-ray flares from Sgr A$^*$}\label{subsec:flares}
Forming the dynamical center of the Milky Way, located at a distance of $\simeq$8~kpc, \sgra\ is the most nearby supermassive black hole (e.g., \citep[][]{reid2004,ghez2008,gillessen2009}). As such it allows for an unparalleled study of how galactic nuclei accrete and supply feedback to their environment. Most remarkably, despite having an estimated mass of $\simeq$$4 \times 10^{6}~\Msun$, the bolometric luminosity output of \sgra\ is only a factor of $\simeq550$ higher than that of the Sun. This is $\simeq$$10^{9}$ times fainter than expected for Eddington-limited accretion onto a black hole of this mass (e.g., \citep[][]{melia2001,genzel2010,morris2012} for reviews). \sgra\ likely feeds off the winds of nearby massive stars (e.g., \citep[][]{coker1997,quataert1999,cuadra2008}), but only $\simeq$1\% of the matter captured at the Bondi radius seems to reach the supermassive black hole \citep[][]{wang2013}. Its puzzling sub-luminous character can be explained if the majority of the matter is ejected and the accretion flow is radiatively inefficient. This may well be the dominant form of accretion onto supermassive black holes throughout the Universe \citep[e.g.,][]{ho1999,nagar2005}, and it is therefore of great interest to understand the fueling process of our Galactic nucleus.

Despite its faint persistent X-ray emission of $L_{\mathrm{X}}$$\simeq$$3\times10^{33}~\lum$ (2--10 keV; \citep[][]{baganoff2001,baganoff2003}), \sgra\ is not completely inactive; roughly daily its X-ray emission flares up by a factor of $\simeq5-150$ for tens of minutes to a few hours. Several tens of such X-ray flares have been detected to date with different satellites (e.g., \citep[][]{baganoff2001,baganoff2003,goldwurm2003,belanger2005,porquet2003,porquet08,trap2011,nowak2012,degenaar2013_sgra,neilsen2013,barriere2014,mossoux2015}).\footnote{Similar (often associated) flaring activity is observed at near-infrared wavelengths (see e.g., \citep[][]{witzel2012} and references therein).} Most have an intensity of $L_{\mathrm{X}}$$\lesssim$$10^{35}~\lum$ in the 2--10 keV band; such weak X-ray flares occur approximately once every day \citep[][]{neilsen2013}. On a few occasions, however, bright flares with $L_{\mathrm{X}}$$\simeq$$(1-5) \times 10^{35}~\lum$ have been detected \citep[][]{baganoff2001,porquet2003,porquet08,nowak2012,degenaar2013_sgra,barriere2014,reynolds2014}. The overall duration and short-timescale variability of the X-ray flares suggests that the emission originates close to the black hole, within $\simeq$10$~R_s$ (where $R_s$$=$$2GM/c^2$ is the Schwarzschild radius, e.g., \citep[][]{baganoff2001,porquet2003,barriere2014,mossoux2015}). 
These events therefore provide excellent means to investigate the inner accretion flow, offering a new view of the feeding processes of \sgra. Different mechanisms have been proposed to explain the flares, including magnetic reconnection and particle acceleration processes (e.g., \citep[][]{markoff2001,yuan2003,liu2004,liu2006}), as well as the infall of gas clumps or disruption of small bodies such as asteroids or comets (e.g., \citep[][]{cadez2006,tagger2006}). Constraining the repetition rate and spectral properties of X-ray flares is an important aspect of understanding their origin (e.g., \citep[][]{porquet08,trap2010}). 

\begin{figure}[tb!]
 \begin{center}
\includegraphics[width=9.5cm]{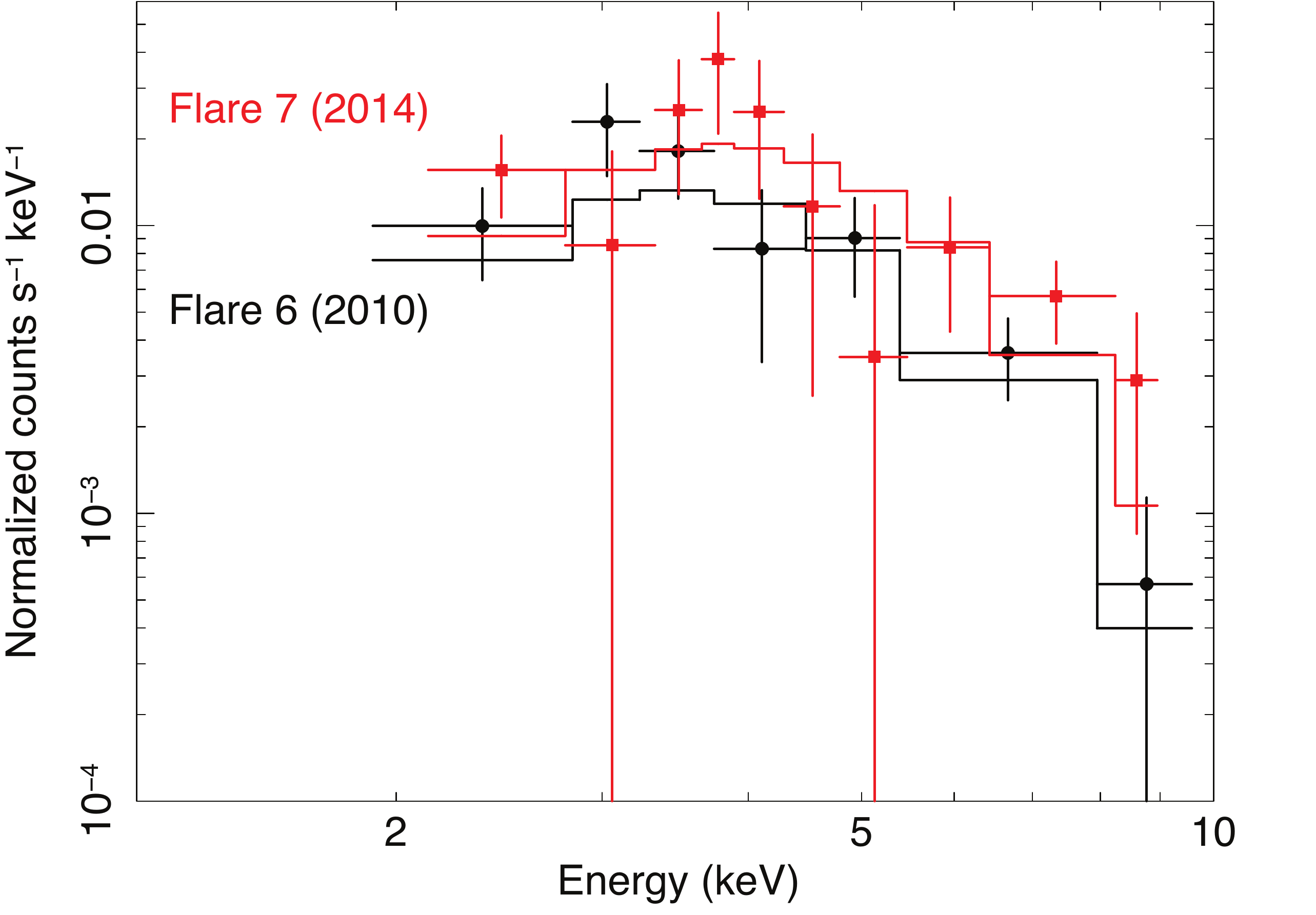}\vspace{-0.5cm}
    \end{center}
\caption[]{{Comparison between the two brightest flares detected from \sgra\ with \swift: the new 2014 X-ray flare (red, squares; \citep[][]{reynolds2014}) and the one observed in 2010 (black, circles; \citep[][]{degenaar2013_sgra}). The solid lines correspond to a fit with an absorbed power-law model, where the hydrogen column density was kept fixed at $N_{\mathrm{H}}=9.1\times10^{22}~\nh$. 
}}
\vspace{-0.4cm}
 \label{fig:flares}
\end{figure}

\swift's X-ray monitoring campaign of the Galactic center has proven to be a powerful tool to catch and study X-ray flares from \sgra; out of the 13 bright X-ray flares ($L_{\mathrm{X}}$$\gtrsim$$1 \times 10^{35}~\lum$) reported to date, seven were detected by \swift. The spatial resolution of the XRT does not allow us to separate \sgra\ from its dense environment; in a $10''$ aperture centered at the radio position of the supermassive black hole we detect a continuum emission of $L_{\mathrm{X}}$$\simeq$$2 \times 10^{34}~\lum$ (2--10 keV). This is a factor of $\simeq$10 higher than the quiescent emission of \sgra\ and is dominated by the X-ray emission of diffuse structures and a number of faint X-ray point sources \citep[][]{baganoff2003,degenaar2013_sgra}. Nevertheless, the XRT can  pick up X-ray flares since these can be an order of magnitude brighter. Indeed, in an initial study using all XRT-PC data obtained in 2006--2011, a total of six flares with intensities of $L_{\mathrm{X}}$$\simeq$$(1-2) \times 10^{35}~\lum$ were identified \citep[][]{degenaar2013_sgra}. This is illustrated in Figure~\ref{fig:sgralc}, which shows the long-term XRT count rate light curve of \sgra. The background-subtracted X-ray spectrum of the flare that collected the most photons (number 6, 2010 June 12), is shown in Figure~\ref{fig:flares}. All \swift-detected flares are positionally coincident with \sgra\ and their duration, spectra, and peak luminosities are similar to flares detected from the supermassive black hole with \chan\ and \xmm. None of the other X-ray point sources located within the XRT extraction region (e.g., accreting white dwarfs and X-ray binaries) have ever been seen to display such short (hours) X-ray variability peaking at $L_{\mathrm{X}}$$\simeq$$10^{35}~\lum$. Therefore, it is highly likely that all six events were X-ray flares from \sgra\ \citep[][]{degenaar2013_sgra}.

The \swift\ study of \sgra\ more than doubled the number of bright X-ray flares observed at that time. Furthermore, the unprecedented 6-year long baseline of daily--weekly observations allowed for an estimate of the occurrence rate of bright X-ray flares; we found \sgra\ fires off one such event every $\simeq$5--10 days \citep[][]{degenaar2013_sgra}. Finally, with such a large number of X-ray flares detected by a single instrument, we were able to perform a comparative study of their spectral properties unbiased by instrumental effects that are present when flares detected by different instruments are compared. Our analysis suggested that despite their similar intensities, one of the flares (the sixth, shown in Figure~\ref{fig:flares}) possibly had a different spectral shape than the other five. Although this was a stand-alone and low-significance result at the time, similar findings were recently reported from comparing two bright X-ray flares detected with \nustar\ \citep[][]{barriere2014}. Taken together, this may indicate that X-ray flares can have different spectral shapes, which may give further insight into their nature. For instance different spectral shapes can naturally be explained if the X-ray flares are related to magnetic reconnection events, which create turbulence and stochastic acceleration of electrons, leading to a range of particle distribution slopes (e.g., \citep[][]{liu2004,zharkova2011}). Nevertheless, analysis of 39 X-ray flares caught by \chan, including one bright, did not reveal spectral differences between individual flares \citep[][]{neilsen2013}. Continued investigation is therefore warranted.

In 2013 April, \swift\ lost its view of \sgra\ due to sudden activity of the nearby magnetar \magnetar, with an angular separation of only $2.4''$ and a 2--10 keV X-ray luminosity a factor $\simeq$100 higher than the quiescent emission of the supermassive black hole (see Section~\ref{subsec:magnetar}). This put \swift's detection of X-ray flares temporarily on hold. However, the X-ray emission of the magnetar steadily faded and became faint enough for the XRT to pick up a bright X-ray flare from \sgra\ again in 2014 September \citep[][]{reynolds2014,degenaar2014_gcflare}.

 \begin{figure}[t!]
 \begin{center}
\includegraphics[width=8.0cm]{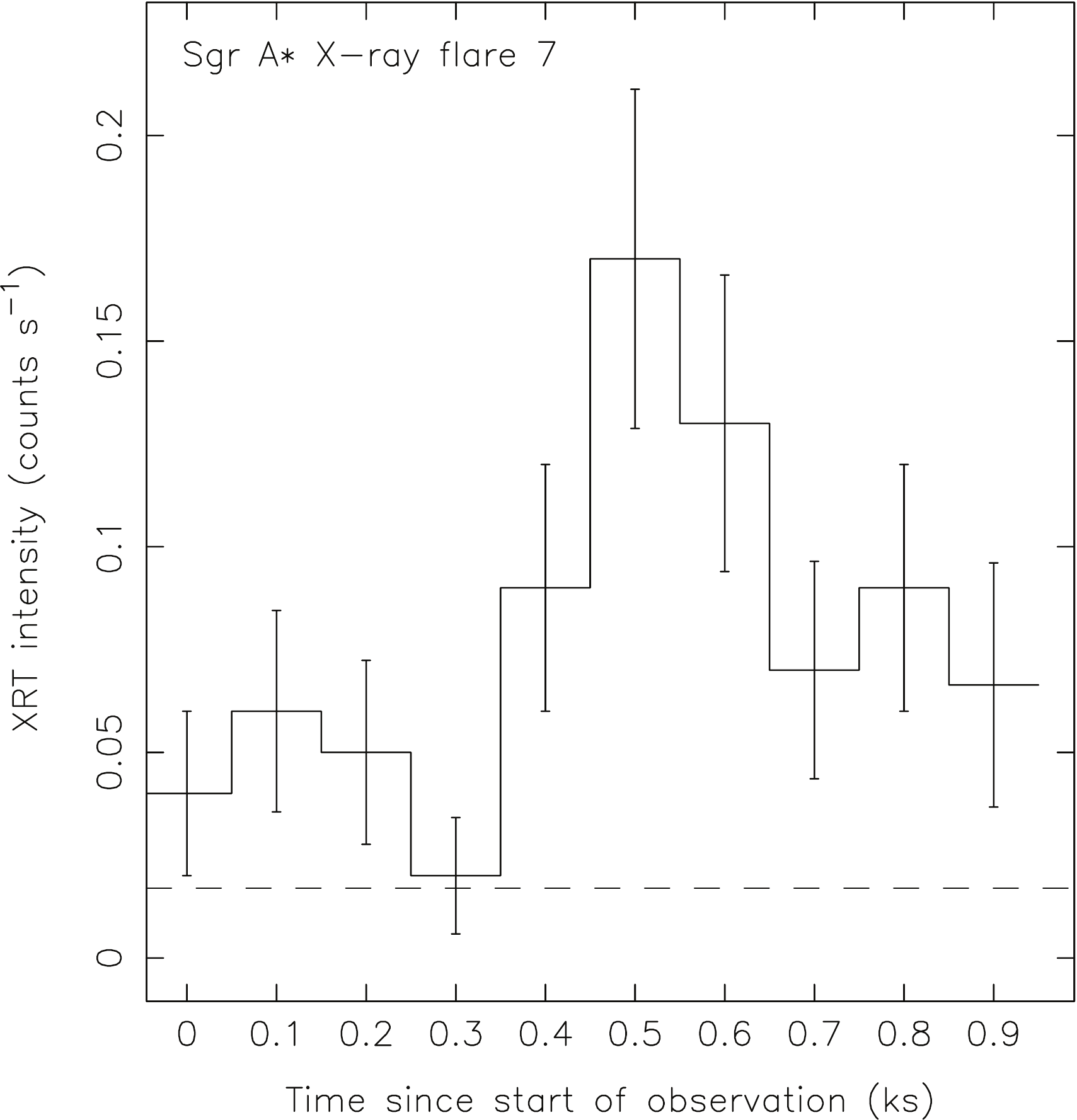}\vspace{-0.5cm}
    \end{center}
\caption[]{{Light curve of the 2014 \swift\ X-ray flare observation, at 100-s resolution. The dashed horizontal line indicates the average count rate observed in the preceding and subsequent observation. 
}}
\vspace{-0.4cm}
 \label{fig:flarelc}
\end{figure}

\subsubsection{The seventh X-ray flare detected in 2014 September}\label{subsubsection:newflare}
\citet{reynolds2014} reported the detection of a new X-ray flare from \sgra, which occurred during a $\simeq$1-ks observation that started on 2014 September 9 at 11:41 UT (obsID 91906150). Figure~\ref{fig:flarelc} shows the XRT light curve of this observation. The count rate varied between $\simeq$(2--20)$\times10^{-2}~\cnts$, with a mean value of $(8\pm1)\times10^{-2}~\cnts$. As can be seen in Figure~\ref{fig:flarelc}, this is significantly higher than the count rate detected at this position in the preceding observation  on September 8 at 04:01--05:49 UT, i.e., $\simeq$30 hr earlier (obsID 91906149; $(1.8\pm0.4)\times10^{-2}~\cnts$), and the subsequent pointing on September at 10 06:52--07:10 UT, i.e., $\simeq$19 hours after the flare detection (obsID 91906151, $(1.6\pm0.4)\times10^{-2}~\cnts$; \citep[][]{degenaar2014_gcflare}). The event thus had a duration of $\gtrsim$16~min, but $\lesssim$49~hr. The timescale, intensity and spectrum (see below), are similar to X-ray flares detected previously from \sgra\ with \swift, strongly suggesting that this event too was an X-ray flare from the supermassive black hole \citep[][]{reynolds2014,degenaar2014_gcflare}.

Extracting a single spectrum from a $10''$ circular region yielded 79 counts, i.e., sufficient to carry out basic spectral fitting. We used the same data reduction and analysis approach as outlined in \citet{degenaar2013_sgra}. However, given that the magnetar \magnetar\ was likely still contributing to the local background of \sgra\ in 2014 (see Section~\ref{subsec:magnetar}), we cannot use the same background model as used for previous X-ray flares. To model the background for the 2014 flare we therefore extracted a spectrum using the 10 observations preceding the detection (obsIDs 91906140--149) and 10 observations following it (obsIDs 91906151--160). This averaged continuum background spectrum can be described by an absorbed power-law model (where we fixed $N_{\mathrm{H}}=9.1\times10^{22}~\nh$ as found from analyzing the non-flaring emission in 2006--2011; \citep[][]{degenaar2013_sgra}) with $\Gamma=1.9\pm0.2$, and a 2--10 keV (absorbed) flux of $F^{\mathrm{abs}}_{\mathrm{X}} = (2.2 \pm 0.1) \times 10^{-12}~\flux$. We then fitted the flare spectrum including background to a double power-law model, with the parameters for the continuum fixed to these values. The results are given in Table~\ref{tab:newflare}. 

The estimated luminosity of $L_{\mathrm{X}} \simeq 1\times10^{35}~\lum$ for the 2014 flare is similar to the brightest one detected previously with \swift\ on 2010 June 12 (see Figure~\ref{fig:sgralc}). This makes it interesting to model the two events together, to see if there are any spectral differences as hinted previously \citep[][]{degenaar2013_sgra}. Given the different background emission for the two flares (see above)\footnote{\magnetar\ very likely resided in quiescence in 2010, implying an X-ray luminosity of $L_{\mathrm{X}} \lesssim 10^{32}~\lum$ \citep[][]{mori2013}.}, we opted for both flares to subtract a spectrum obtained from the preceding observation, rather than modeling the background emission. This way we obtained  $\Gamma=1.9 \pm 0.9$ for 2010 and $\Gamma = 1.1 \pm 1.1$ for 2014 (with $N_{\mathrm{H}}=9.1\times10^{22}~\nh$ fixed). The spectral indices are consistent within the 90\% confidence errors, i.e., we find no spectral differences between these two flares.

\begin{table}[t!]
\caption{Properties of the seventh bright X-ray flare from \sgra.}
\begin{threeparttable}
\begin{tabular*}{0.48\textwidth}{@{\extracolsep{\fill}}lc}
\hline
Parameter & Value \\
\hline
ObsID \dotfill & 91906150  \\ 
Date \dotfill & 2014 September 09  \\ 
Obs start time (UTC; hh:mm:ss.s) \dotfill & 11:41:29.5   \\ 
Flare duration (hr) \dotfill &  $0.27 \lesssim t \lesssim 49$ \\ 
Rate ($\cnts$) \dotfill &  $(8\pm1)\times10^{-2}$ \\ 
HR  \dotfill & $1.51 \pm 0.23$  \\ 
$\Gamma$  \dotfill & $0.9 \pm 0.8$  \\ 
Cstat/dof  \dotfill & 46.46/69  \\ 
$F_{\mathrm{X}}^{\mathrm{abs}}$ ($\flux$)  \dotfill & $(2.6 \pm 0.7) \times 10^{-11}$ \\  
$L_{\mathrm{X}}$ ($\lum$)  \dotfill & $(1.4 \pm 0.4) \times 10^{35}$ \\ 
\hline
\end{tabular*}
\label{tab:newflare}
\begin{tablenotes}
\item[] Note. -- The spectrum of the flare, including background, was fitted with a double power-law model where the parameters for the power-law describing the continuum emission were fixed (see text). 
The hardness HR gives the ratio of counts in the 2--4 and 4--10 keV bands. 
$F_{\mathrm{X}}^{\mathrm{abs}}$ is the {\it absorbed} 2--10 keV flux ($N_{\mathrm{H}}=9.1\times10^{22}~\nh$ was fixed), whereas $L_{\mathrm{X}}$ gives the 2--10 keV luminosity corrected for absorption, assuming $D=8$~kpc. Errors are 90\% confidence levels.
\end{tablenotes}
\end{threeparttable}
\end{table}

\subsubsection{Possible future re-activation of \sgra}\label{subsubsection:reactivate}
The detection of Fermi bubbles, relic jets, and time-variable fluorescent light echoes from giant molecular clouds in the Galactic center, suggest that \sgra\ may have been much more active and orders of magnitude brighter in the (recent) past (e.g., (see e.g., \citep[][]{ponti2012,yusefzadeh2012,li2013,ryu2013} for recent studies and reviews). Therefore it is plausible that our supermassive black hole reactivates again in the future. This idea got revived when in 2012 the discovery was announced that a gaseous object denoted as `G2' was on a collision course with \sgra, passing as close as $few\times10^{3}~R_s$ (e.g.,\citep[][]{gillessen2012,eckart2014_g2,ghez2014_g2,pfuhl2015}). This is well within the Bondi accretion radius of $\simeq10^{5}~R_s$, suggesting that G2 could interact with the ambient hot medium, perhaps creating a bowshock (e.g., \citep[][]{sadowski2013}). Its estimated mass of a few Earth masses is comparable to that present in the accretion flow around \sgra\ \citep[e.g.,][]{yuan2003}. Therefore, if G2 would become disrupted due tidal forces and a significant fraction of the shredded gas would accrete onto the supermassive black hole, that could lead to a significant brightening unfolding over the next decade (e.g.,\citep[][]{anninos2012,gillessen2012_2,schartmann2012}). Ever since this discovery, however, there has been considerable debate about the nature and origin of G2 - whether it is a pure gas cloud or whether there is a hidden central object that could keep it gravitationally bound - resulting in a wide range of predictions how the activity of \sgra\ would be affected (e.g., \citep[][]{burkert2012,schartmann2012,meyer2012,miralda2012,phifer2013,scoville2013,decolle2014,guillochon2014}). 

The prospective interaction between G2 and \sgra\ spurred great interest, because X-ray brightening of the supermassive black hole could give more insight into the physics of Bondi accretion. Dedicated monitoring and target-of-opportunity programs have therefore been set up accordingly, covering almost the entire electromagnetic spectrum. \swift\ plays a central role in this effort since enhanced accretion activity of \sgra\ should manifest itself in the X-ray band. Given that the \swift/XRT monitoring program is a factor $\simeq$10--100 more sensitive than daily scans from \maxi\ and \swift/BAT, but can still target the Galactic center almost every day, it has been widely recognized that this \swift\ program would serve as an important trigger for other observatories in case the interaction with G2 were to affect the supermassive black hole. In this respect, having mapped out the years-long X-ray activity of \sgra\ with \swift\ provides an important calibration point, allowing us to detect any possible changes in its accretion activity (e.g., not only an increase in its persistent emission, but also in its X-ray flaring activity; \citep[][]{degenaar2013_sgra}). 

Deep near-infrared observations have shown that G2 completed its closest approach in mid-2014, although it is debated whether there are signs of disruption \citep[][]{pfuhl2015,witzel2014}. At the time of writing, no enhanced activity of \sgra\ has been detected in the X-ray band, nor at other wavelengths \citep[][]{crumley2013,sadowski2013,chandler2014,haggard2014,hora2014,tsuboi2015}. Nevertheless, transporting gas out from $\simeq10^{3}~R_s$ to the inner accretion flow would take a viscous time scale, which is estimated to be years for \sgra\ (e.g., \citep[][]{burkert2012,schartmann2012,moscibrodzka2012}). Continued, frequent monitoring in X-ray band is therefore highly desired to keep on watch for any possible enhancement in the activity of the supermassive black hole. Only \swift\ can accommodate high-cadence, sensitive X-ray monitoring to explore the interaction with G2 and its future consequences.

\subsection{The Galactic center magnetar \magnetar}\label{subsec:magnetar}
On 2013 April 24, \swift\ detected sustained X-ray activity at the position on \sgra\ \citep[][]{degenaar2013_flare}, which was initially thought to be linked to the anticipated interaction between the supermassive black hole and G2 (Section~\ref{subsubsection:reactivate}). However, a very brief and highly energetic burst seen by \swift/BAT and 3.76-s coherent X-ray pulsations detected by \nustar\ revealed that instead of \sgra\ it was a nearby (angular separation of only $2.4''$; \citep[][]{rea2013}), highly-magnetized neutron star that had suddenly become X-ray active: \magnetar\ \citep[][]{mori2013,kennea2013}. This has been an interesting discovery because the neutron star may be in a bound orbit around the supermassive black hole \citep[][]{rea2013,bower2015}, and its long-term flux evolution can give insight into the physics of neutron star magnetospheres and the composition/structure of their interior \citep[][]{kaspi2014,cotizelati2015}. Moreover, \magnetar\ was also detected as a radio pulsar \cite[][]{shannon2013}, and its polarized radio emission provided a measurement of the magnetic field near \sgra\ \citep[][]{eatough2013}. Finally, its discovery provided new strategies for finding other radio pulsars near \sgra\ that could probe space-time deformations around the supermassive black hole \citep[][]{bower2014}.

Quite remarkably, the X-ray outburst of \magnetar\ faded much more slowly than typically seen for transient magnetars \citep[][]{kennea2013,kaspi2014,cotizelati2015}; the source continued to be detected in the 2--10 keV band over 1.5 years after it switched on (see Figure~\ref{fig:sgralc}). Fitting the XRT light curve of \magnetar\ to a simple exponential decay yields an e-folding time of $\tau$$=$$152\pm6$~days (with a normalization of $(8\pm1)\times10^{-2}~\cnts$), suggesting that it should fade into the local X-ray background roughly $800$~days after its onset, which would correspond to mid-2015. Indeed, in a $10''$ extraction region around \sgra\ the average count rate in the last 10 observations of 2014 (October 21--November 2) was $(1.8\pm0.1)\times10^{-2}~\cnts$, i.e., higher than the average intensity of $(1.1\pm0.1)\times10^{-2}~\cnts$ in 2006--2011 \citep[][]{degenaar2013_sgra}. Nevertheless, by late 2014 the magnetar had already faded sufficiently to allow the detection of X-ray flares from \sgra\ with the XRT again (Section~\ref{subsubsection:newflare}).

\begin{table*}[tb!]
\begin{center}
\caption{Transient X-ray binaries covered by the \swift\ Galactic center monitoring campaign.}
\begin{threeparttable}
\begin{tabular*}{0.99\textwidth}{@{\extracolsep{\fill}}lcccccc}
\hline
Source name & Year  & $L_{\mathrm{X}}$ & $L_{\mathrm{X}}^{\mathrm{peak}}$& $t_{\mathrm{ob}}$ & Classification & Reference \\
& &  \multicolumn{2}{c}{($10^{35}~\lum$)} & (weeks) &  & \\
\hline
1. \ascabron & 2006 & $4.1 \times 10^{35}$ & $9.2 \times 10^{35}$& $>16$ & Neutron star LMXB & \citep[][]{degenaar09_gc} \\
& 2007--2008 & $1.1\times 10^{36}$ & $5.1\times 10^{36}$& $>80$ & &  \citep[][]{degenaar09_gc,degenaar2010_gc} \\
& 2010 & $2.7 \times 10^{35}$ & $5.6 \times 10^{35}$ & 20--34 & &  \citep[][]{degenaar2014_gctransients} \\
& 2013--2015$^{*}$ & $2.5\times 10^{36}$ & $4.7\times 10^{36}$ & $>84$  & & \citep[][]{degenaar2014_gctransients} \\
2. \brontwee & 2006 & $1.6 \times 10^{34}$ & $4.0 \times 10^{34}$& $>8$ & VFXB & \citep[][]{degenaar09_gc} \\
& 2008 & $1.1 \times 10^{34}$ & $3.0 \times 10^{34}$& $>12$ &  & \citep[][]{degenaar2010_gc} \\
3. \brondrie & 2006 & $9.6 \times 10^{34}$ & $2.3 \times 10^{35}$& $2$  & VFXB & \citep[][]{degenaar09_gc} \\
& 2013 & $1.5\times 10^{35}$ & $4.0\times 10^{35}$& $4$ & & \citep[][]{koch2014} \\
4. \bronvier & 2006 & $6.0 \times 10^{34}$ & $1.5 \times 10^{35}$& $2$ & VFXB & \citep[][]{degenaar09_gc} \\
5. \bronvijf & 2006 & $1.2 \times 10^{34}$ & $5.0 \times 10^{34}$& $5$ & VFXB & \citep[][]{degenaar09_gc} \\
6. \grsbron & 2006 & $3.0 \times 10^{34}$ & $5.0 \times 10^{34}$& $1$ & Neutron star LMXB & \citep[][]{degenaar09_gc} \\
 & 2007 & $1.1\times 10^{36}$ & $2.0\times 10^{36}$& $>13$ &  & \citep[][]{degenaar09_gc} \\
 & 2009 & $1.8\times 10^{36}$ & $1.3\times 10^{37}$& 4--5 &  & \citep[][]{degenaar2010_gc} \\
 & 2010 & $6.3 \times 10^{35}$ & $1.4\times 10^{36}$ & $13$ &  & \citep[][]{degenaar2014_gctransients} \\
& 2013 & $4.4\times 10^{36}$ & $2.3\times 10^{37}$ & $6$ & & \citep[][]{degenaar2014_gctransients} \\
7. \xmmbron & 2007 & $2.6 \times 10^{33}$ & $7.3 \times 10^{33}$& $<12$ & Neutron star LMXB & \citep[][]{degenaar09_gc} \\
& 2008 & $1.7 \times 10^{35}$ & $1.4 \times 10^{36}$& 1--7 &  & \citep[][]{degenaar2010_gc} \\
& 2009 & $1.0 \times 10^{34}$ & $1.1 \times 10^{35}$& $<2$ &  & \citep[][]{degenaar2010_gc} \\
& 2010 & $2.6 \times 10^{35}$ & $1.9 \times 10^{35}$& 1--3 &  & \citep[][]{degenaar2014_xmmsource} \\
& 2012 & $3.4 \times 10^{35}$ & $9.8 \times 10^{35}$& $3$ &  &\citep[][]{degenaar2014_xmmsource} \\
& 2013 & $3.0\times 10^{34}$ & $4.5\times 10^{34}$& 1 &  & this work \\
& 2014 & $1.0\times 10^{34}$ & $1.1\times 10^{34}$& 1--2 &  & this work \\
8. \bronacht & 2009 & $3.8 \times 10^{34}$ & $1.7 \times 10^{35}$& 30--52 & VFXB & \citep[][]{degenaar2010_gc}  \\
9. \bronnegen & 2011 & $5.1 \times 10^{34}$ & $1.1 \times 10^{35}$& 1--2 & VFXB & this work  \\
\hline
\adcbron & 2004--2005 & -- & $1 \times 10^{35}$ & -- &  Black hole LMXB?  & \citep[][]{muno05_apj633,muno05_apj622,porquet05_eclipser} \\
\othercxobron & 1999--2005 & -- & $1 \times 10^{34}$ & -- & VFXB &   \citep{muno05_apj622} \\
1A 1742--289 & 1975 & -- & $7 \times 10^{38}$ & -- &   Black hole LMXB? &   \citep{davies1976,branduardi1976,muno2009}\\
XMMU J174554.4--285456 & 2002 & -- & $8 \times 10^{34}$ & -- &  VFXB &   \citep[][]{muno05_apj622,porquet05} \\
XMM J174544--2913.0 & 2000 & -- & $5 \times 10^{34}$ & -- &  VFXB &   \citep[][]{sakano05} \\
\hline
\end{tabular*}
\label{tab:spec}
\begin{tablenotes}
\item[] Note. -- $^{*}$The latest outburst of \ascabron\ was ongoing at the time of writing (2015 March). This overview is collected from the literature except for the 2011 outburst of \bronnegen\ and the 2013/2014 activity of \xmmbron, which are reported here. Sources are numbered in the order that they appeared active during the campaign. For \xmmbron\ we list only outbursts for which a rise and/or decay could be determined, i.e., we do not list extended periods of low-level activity such as seen from this source in 2008 \citep[][]{degenaar2010_gc} and 2011 \citep[][]{degenaar2014_xmmsource}. $L_{\mathrm{X}}$ gives the average 2--10 keV outburst luminosity and $L_{\mathrm{X}}^{\mathrm{peak}}$ the peak luminosity in that band. We assumed distances of $D=6.7$~kpc for \grsbron\ and $D=6.5$~kpc for \xmmbron\ as inferred from Type-I X-ray burst analysis (\citep[][]{trap2010} and \citep[][]{degenaar2014_xmmsource}, respectively), whereas $D=8$~kpc was used for all other sources. The bottom five are known X-ray transients within the FOV that were not active between 2006--2014; for reference we list their peak $L_{\mathrm{X}}$ reported for previous outbursts.
\vspace{-0.4cm}
\end{tablenotes}
\end{threeparttable}
\end{center}
\end{table*}

\begin{figure*}[tb!]
 \begin{center}
\includegraphics[width=5.9cm]{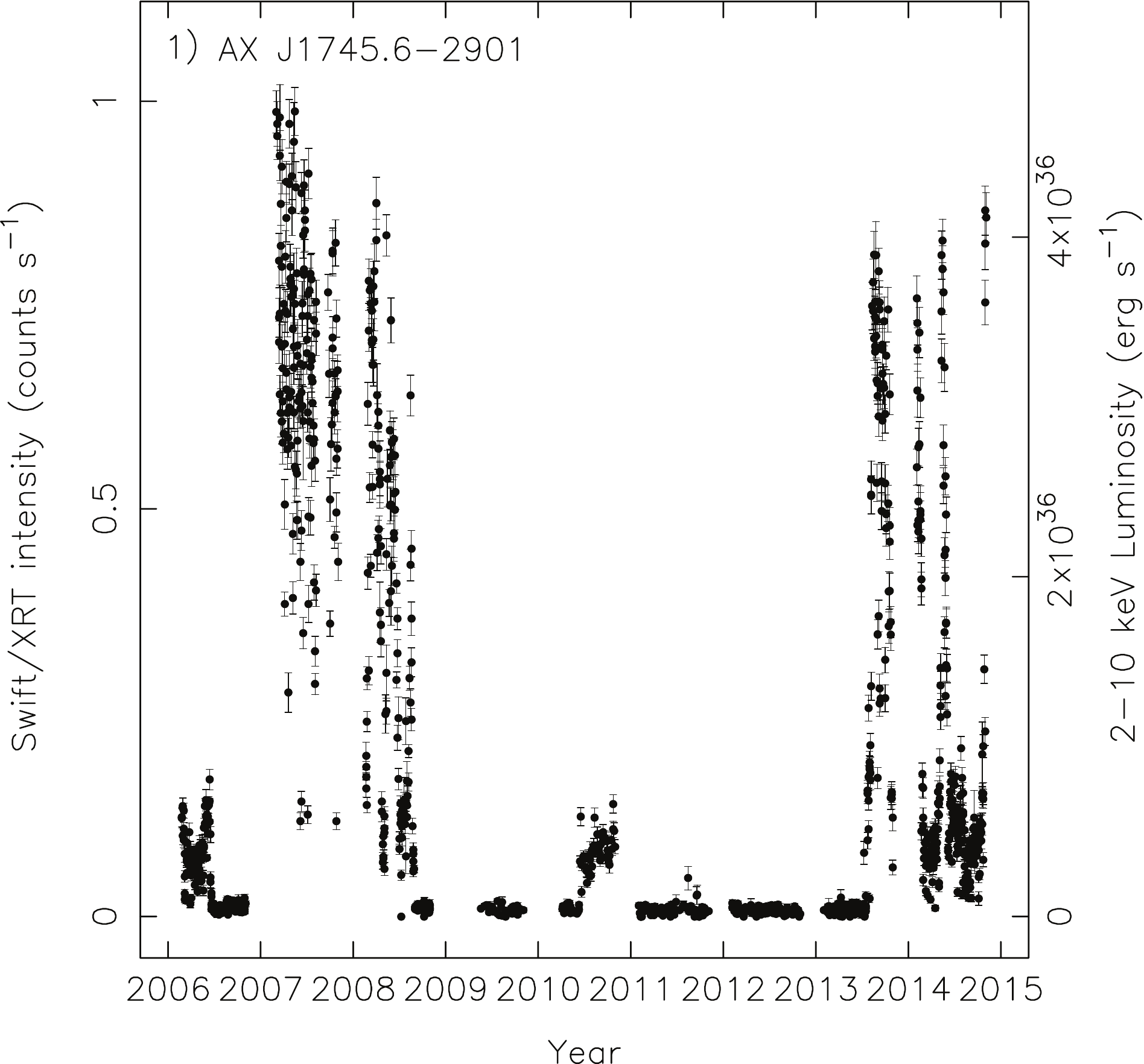}\hspace{+0.2cm}
\includegraphics[width=5.9cm]{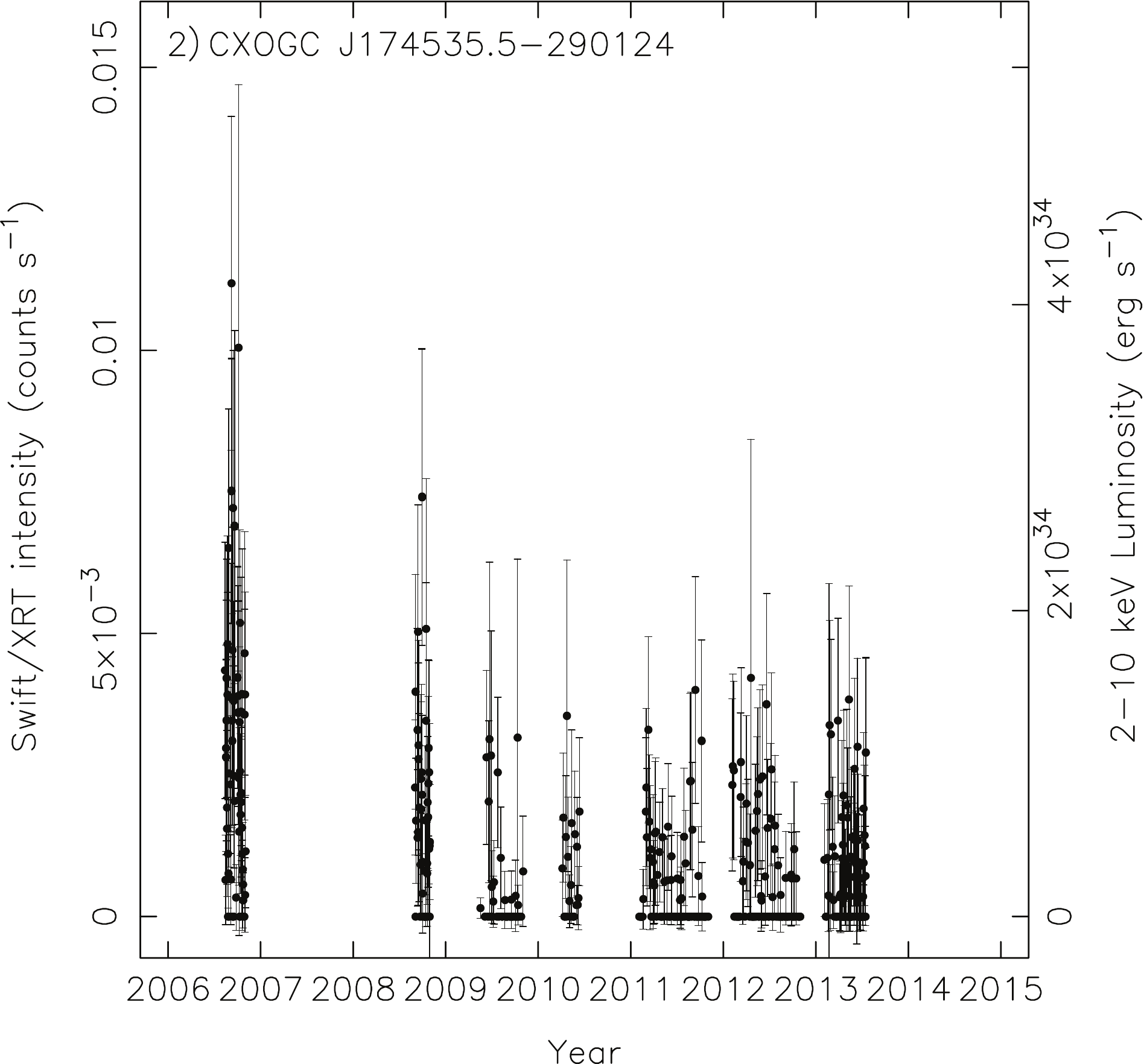}\hspace{+0.2cm}
\includegraphics[width=5.9cm]{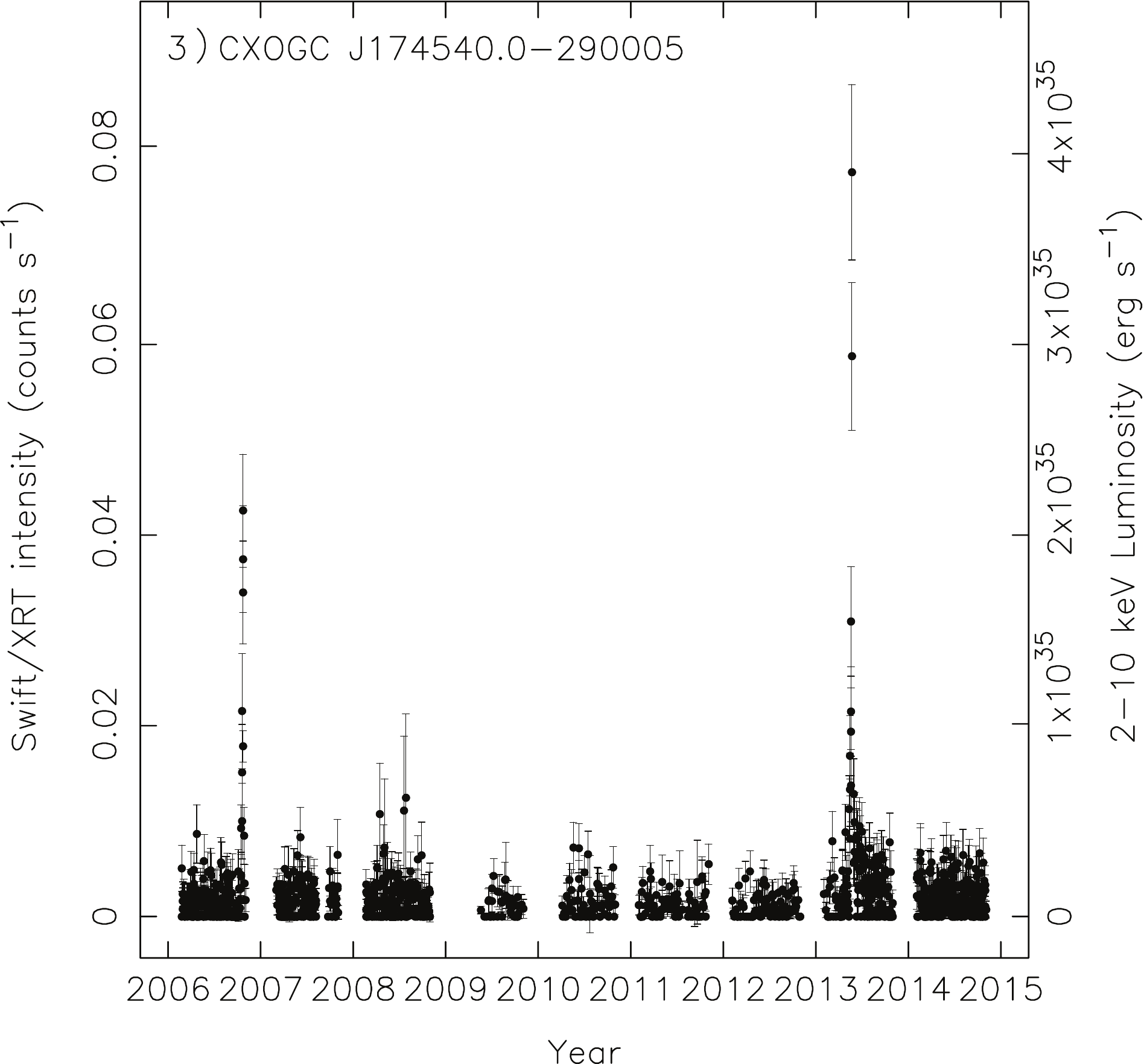}\vspace{+0.2cm}
\includegraphics[width=5.9cm]{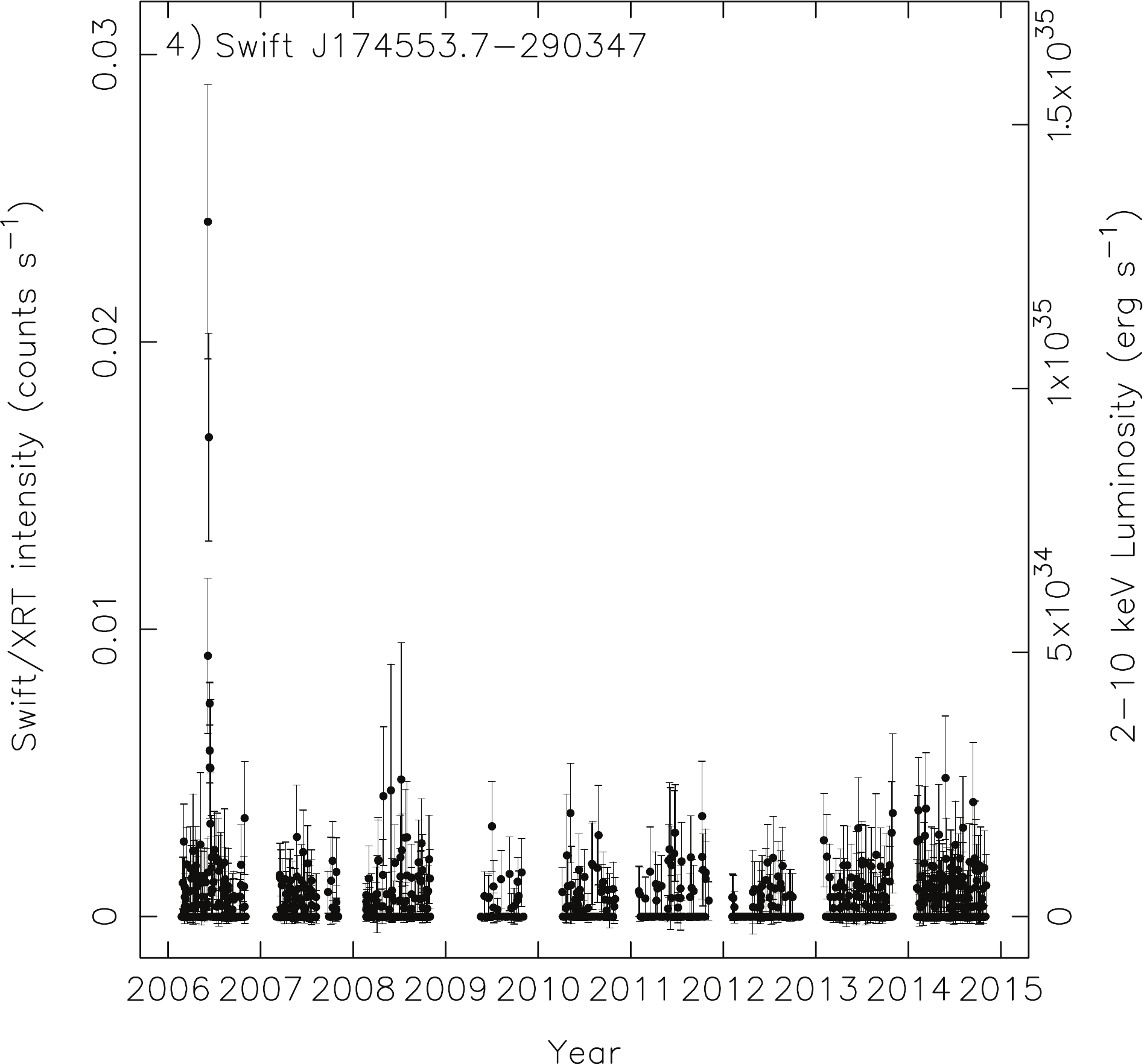}\hspace{+0.2cm}
\includegraphics[width=5.9cm]{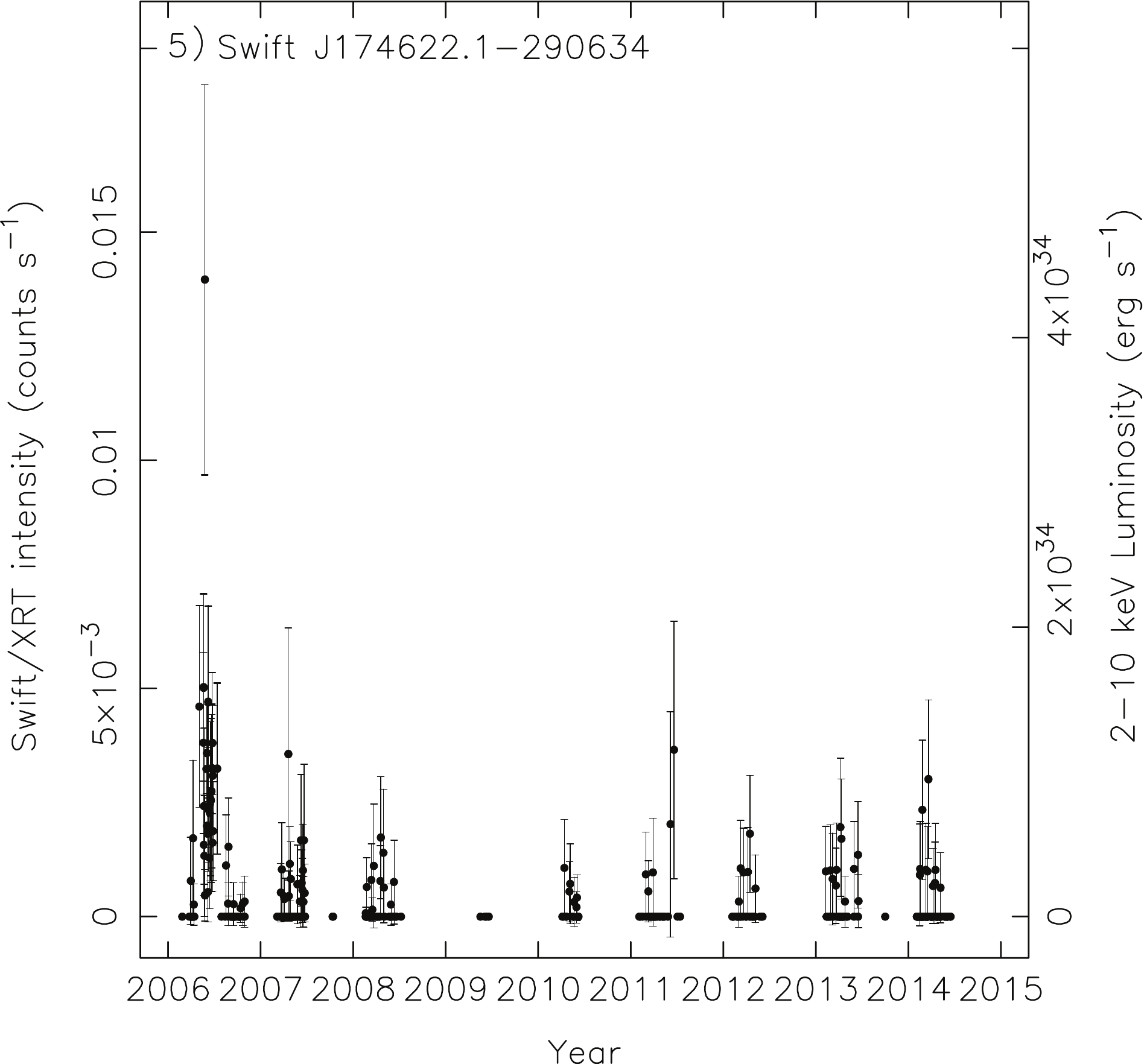}\hspace{+0.2cm}
\includegraphics[width=5.9cm]{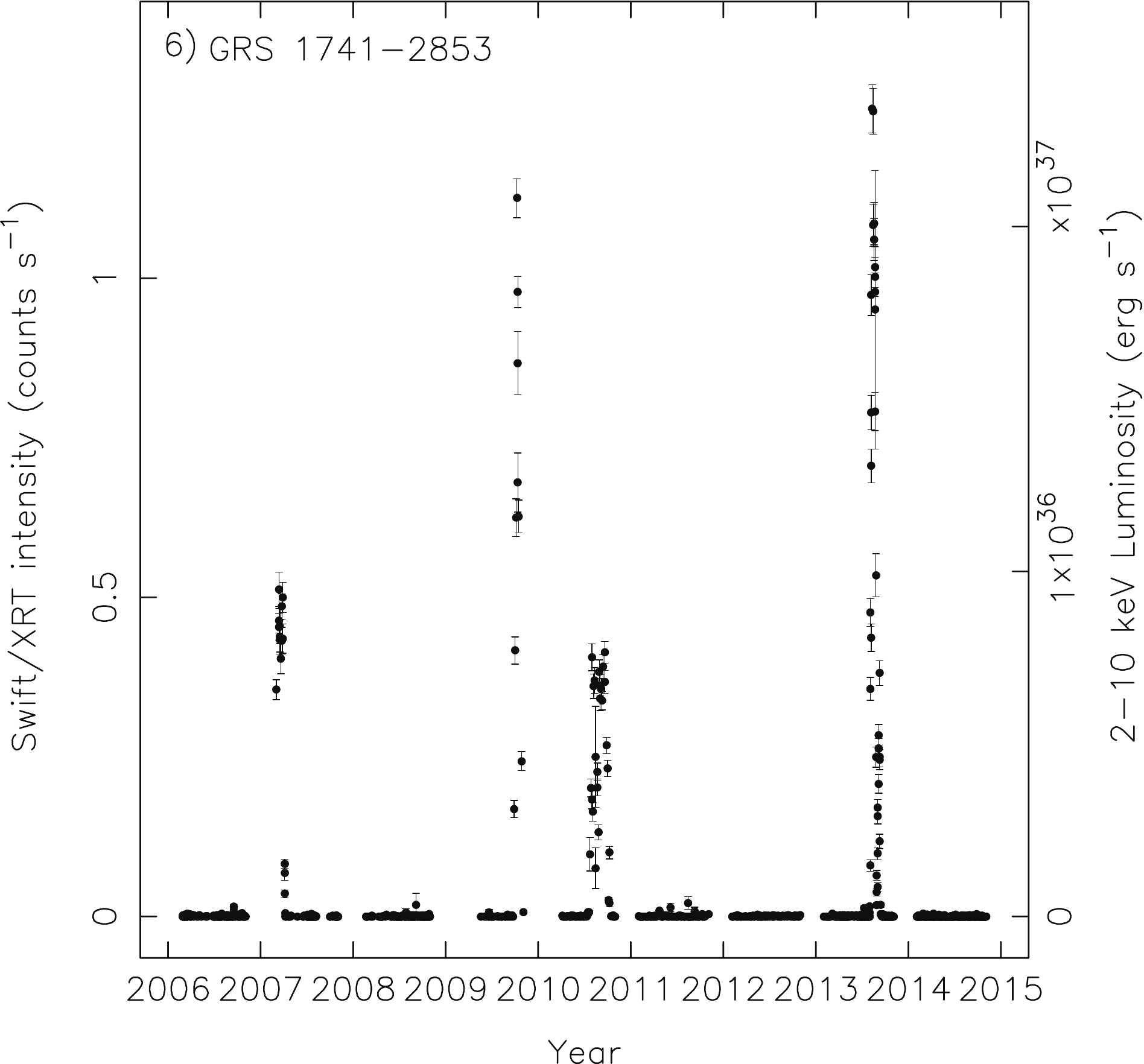}\vspace{+0.2cm}
\includegraphics[width=5.9cm]{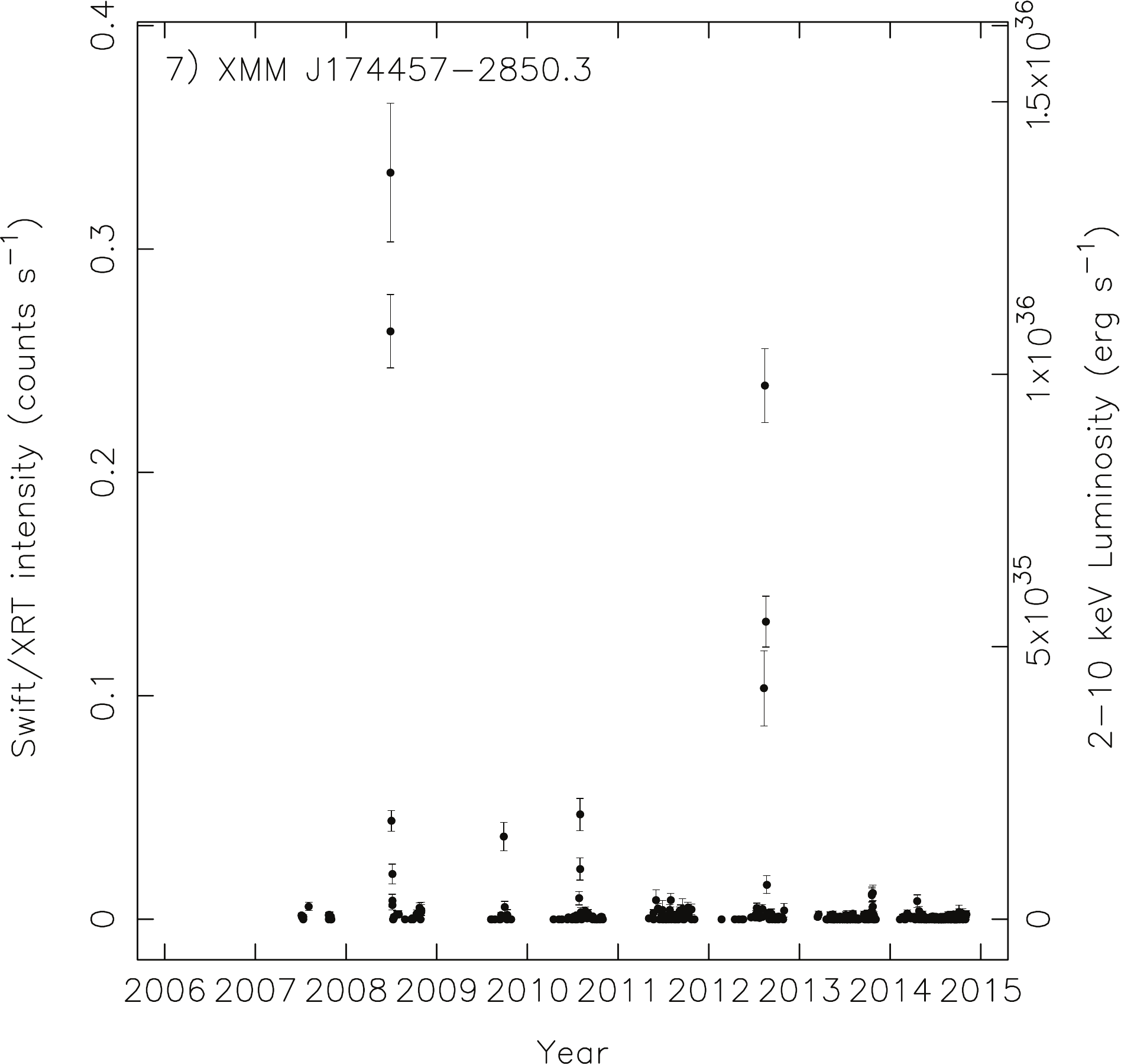}\hspace{+0.2cm}
\includegraphics[width=5.9cm]{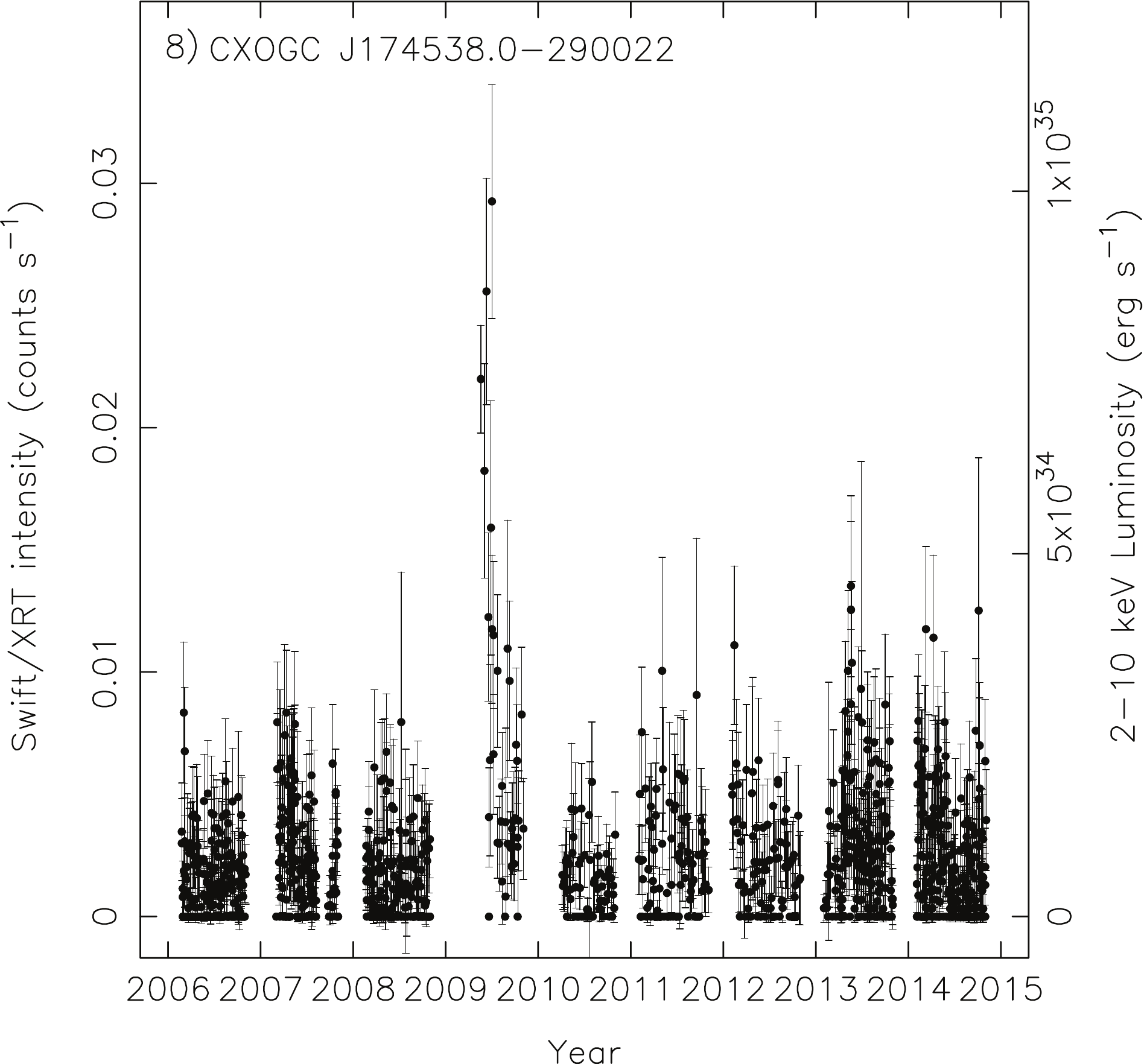}\hspace{+0.2cm}
\includegraphics[width=5.9cm]{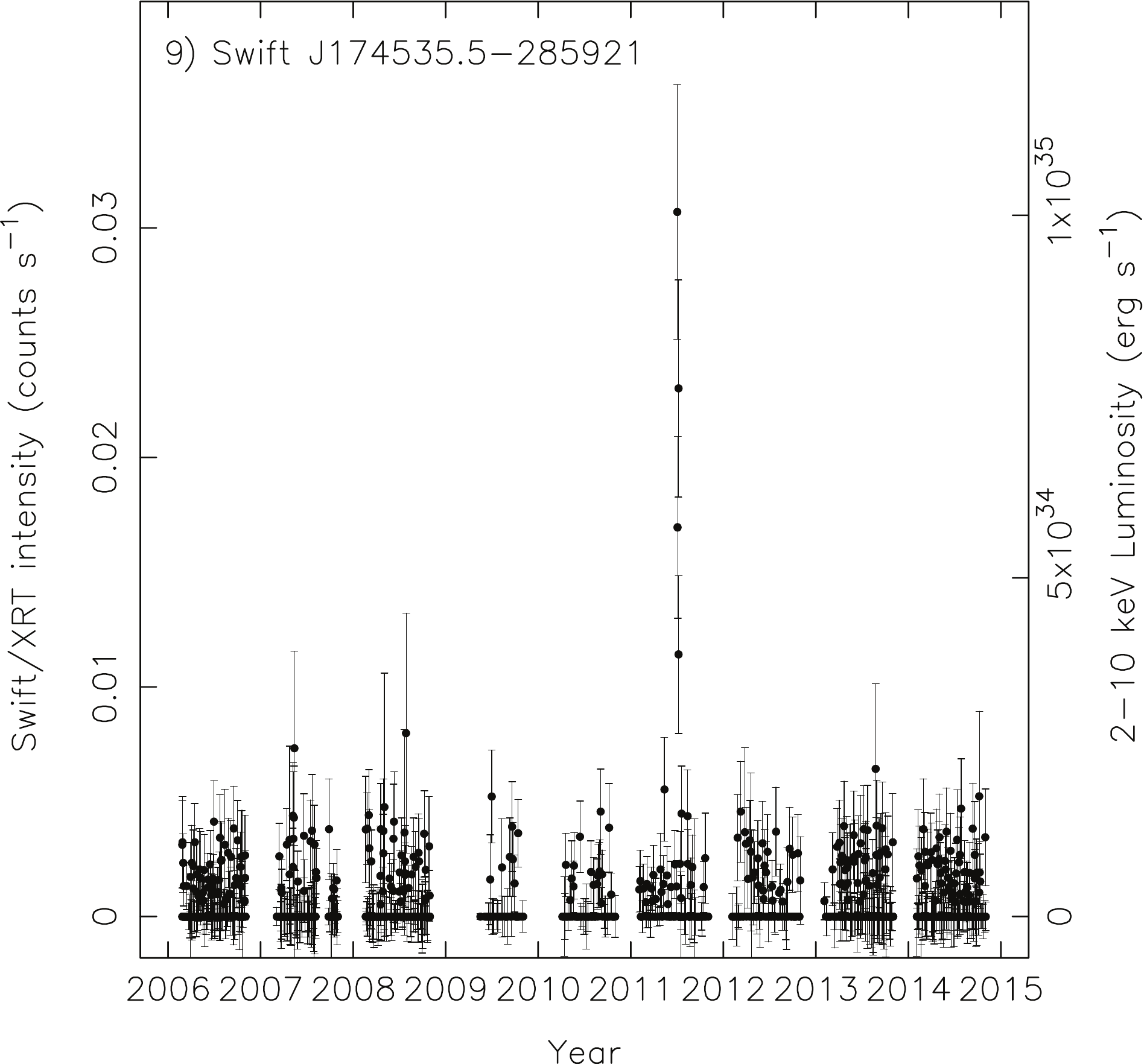}\vspace{-0.5cm}
    \end{center}
\caption[]{{Background corrected long-term 0.3--10 keV \swift/XRT light curves of the 9 transient X-ray binaries that were active in 2006--2014 (PC mode data only, binned per observation). More detailed light curves of selected individual outbursts are presented in Figure~\ref{fig:lowlevel}. Distances were assumed to be $D=8$~kpc, except for \grsbron\ and \xmmbron, for which we adopted the estimates from Type-I X-ray burst analysis ($D=6.7$~kpc and $D=6.5$~kpc, respectively; \citep[][]{trap2010,degenaar2014_xmmsource}).}}
\vspace{-0.4cm}
\label{fig:transientlc}
\end{figure*}

\subsection{The outburst properties of very-faint X-ray binaries}\label{subsec:vf}
In addition to the central supermassive black hole \sgra\ and the Galactic center magnetar \magnetar, the $25' \times 25'$ region covered by \swift\ also includes the position of 14 transient X-ray point sources, 9 of which were seen active in 2006--2014 (Table~\ref{tab:spec}). The locations of these active sources are indicated in Figure~\ref{fig:image}, and their long-term XRT light curves are shown in Figure~\ref{fig:transientlc}. Three of these active sources were previously unknown X-ray transients, uniquely discovered by \swift\ (see Section~\ref{subsec:new}). 

The Galactic center X-ray transients are usually very dim ($L_{\mathrm{X}} \simeq 10^{30-33}~\lum$; \citep[][]{degenaar2012_gc}) and undetected by \swift, but become visible during occasional X-ray outbursts when the 2--10 keV X-ray luminosity rises to $L_{\mathrm{X}} \gtrsim 10^{34}~\lum$ for a duration of $t_{\mathrm{ob}}\gtrsim1$~week (see Table~\ref{tab:spec}). The amplitude, duration and energy output indicates that these X-ray outbursts are powered by accretion onto a neutron star or stellar-mass black hole in an X-ray binary. Indeed three of these sources (\ascabron, \grsbron, and \xmmbron) have displayed Type-I X-ray bursts; bright flashes of X-ray emission caused by thermonuclear runaway on the surface of an accreting neutron star (e.g., \citep[][]{galloway06}). These are generally assumed to have low-mass companion stars and are therefore denoted as low-mass X-ray binaries (LMXBs).

In LMXBs matter is transferred to the compact primary via an accretion disk and transient behavior is ascribed to thermal-viscous instabilities in this disk \citep[e.g.,][for a review]{lasota01}. During the X-ray dim episodes, called quiescence, gas flows from the companion into a cold accretion disk but little of this matter reaches the neutron star/black hole. However, as more material accumulates in the disk the temperature and pressure rise, eventually crossing the threshold for the disk to become hot and ionized. This strongly increases the viscosity and causes matter to rapidly accrete onto the compact primary, resulting in an X-ray outburst. As matter is more rapidly consumed than it is supplied by the companion, the disk becomes depleted and eventually switches back to its cold, quiescent state until a new outburst commences. 

The outburst energetics serve as a probe of the amount of matter that was accreted and the detailed shape of the light curve can be used to determine if this was a large or a small portion of the entire disk (e.g., \citep[][]{king98}). Transient X-ray binaries are often further classified according to their 2--10 keV peak luminosity $L_{\mathrm{X}}^{\mathrm{peak}}$ \citep[][]{wijnands06}. The brightest sources exhibit $L_{\mathrm{X}}^{\mathrm{peak}}$$\simeq$$10^{36-39}~\lum$, but some objects never become brighter than $L_{\mathrm{X}}^{\mathrm{peak}}$$\simeq$$10^{34-36}~\lum$. These are therefore denoted as `very-faint X-ray binaries' (VFXBs; e.g., \citep[][]{muno05_apj622,wijnands06,degenaar09_gc,campana09}). This classification is not necessarily strict, however, since several bright LMXBs also exhibit very-faint outbursts (see Section~\ref{subsec:low}).

As can be seen in Table~\ref{tab:spec}, only four out of 14 X-ray transients covered by the \swift\ campaign have displayed outbursts with $L_{\mathrm{X}}^{\mathrm{peak}}$$>$$10^{36}~\lum$; the majority of sources remain below that level and can thus be classified as VFXBs. Although the VFXBs thus seem to outnumber the brighter LMXBs in the Galactic center region (see also \citep[][]{muno05_apj622,degenaar2012_gc}), their sub-luminous character remains a puzzle. One proposed explanation is that these objects have degenerate (white dwarf) donor stars and short orbital periods, so that their accretion disks are small and hence only little material is accreted over an outburst (e.g., \citep[][]{king_wijn06,zand07,heinke2014_gc}). Another possibility is that these neutron stars have relatively strong magnetic fields that choke the accretion flow (e.g., \citep[][]{wijnands2008_MC,patruno2010_vf,degenaar2014_xmmsource,heinke2014_gc}), or are feeding off the wind of their companion rather than from a disk \citep[][]{pfahl2002,degenaar2010_gc,maccarone2013}. It is also possible that the VFXBs are just bright X-ray binaries that exhibit low-level accretion activity and for which a bright outburst has not been observed yet (e.g., \citep[][]{wijnands2012,heinke2014_gc}, see also Section~\ref{subsec:low}).

The \swift\ Galactic center monitoring program has not only been important to discover new VFXBs (see Section~\ref{subsec:new}), but it also allowed for the first detailed, systematic study of the outburst energetics and recurrence time of these peculiar objects. For instance, the mass-accretion rate averaged over thousands of years, $\langle \dot{M}_{\mathrm{long}} \rangle$, is an important parameter for understanding the evolution of LMXBs. The excellent, long baseline provided by the \swift\ program has allowed to determine the duration ($t_{\mathrm{ob}}$), recurrence time ($t_{\mathrm{rec}}$), and mass-accretion rate of the outbursts of VFXBs ($\dot{M}_{\mathrm{ob}}$), which can then be used to estimate $\langle \dot{M}_{\mathrm{long}} \rangle$ (\citep[][]{degenaar09_gc,degenaar2010_gc,koch2014}, see also Section~\ref{subsec:new}). The resulting low mass-accretion rates  suggest that some VFXBs may indeed have white dwarf companions and hence small orbits/disks (see also \citep[][]{heinke2014_gc}). 

Furthermore, the \swift\ program has provided the first detailed light curves of the (often short) outbursts of VFXBs, which can be compared to accretion disk models to gain information about their disk size \citep[][]{heinke2014_gc}. Studying the light curves of three outbursts of two different sources (\brondrie\ and \xmmbron) revealed that these sources likely have orbital periods of order $\simeq$1~hr and thus indeed have small accretion disks. Such light curve modeling also allows to distinguish between true VFXBs and bright X-ray binaries showing low-level accretion activity (see also Section~\ref{subsec:low}): In the former case the entire (albeit small) disk would be accreted and result in an exponential outburst decay followed by a linear trend, whereas in the latter case only a portion of the disk would be accreted and the resulting decay  would only show a linear trend \citep[][]{heinke2014_gc}. Importantly, this study demonstrated that daily, sensitive X-ray observations are instrumental; less dense sampling does not allow for good constraints on the outburst decays so that no conclusions can be drawn on the disk size. This study of VFXB light curves has not been possible prior to \swift, because all-sky monitors lack the required sensitivity, and other pointed telescopes are not flexible enough to obtain dense sampling of the often short ($<$1 month) outbursts. It is of note that \swift\ also uncovered sustained long periods of low-level accretion activity in one particular source (\xmmbron), which may point toward the magnetic field of the neutron star choking the accretion flow (see Section~\ref{subsec:trans}).

\subsection{Newly discovered (candidate) transient X-ray binaries}\label{subsec:new}
Three of the sources that exhibited an outburst during the \swift\ campaign had previously been identified as dim, non-variable X-ray sources in a deep \chan\ survey \citep[][]{muno2009}, but were not known to be transient; \bronvier, \bronvijf\ \citep[][]{degenaar09_gc}, and \bronnegen\ \citep[][]{degenaar2011_newtransient}. All three exhibited very brief ($\lesssim$5~weeks), and very faint ($L_{\mathrm{X}}\lesssim10^{35}~\lum$) outbursts that are easily missed and have only been detected by the advent of \swift's frequent, sensitive X-ray observations (see Figure~\ref{fig:transientlc} and Table~\ref{tab:spec}). The detailed properties of the newly discovered transients \bronvier\ and \bronvijf\  (both active in 2006) were  reported in \citet{degenaar09_gc}, but we discuss the 2011 discovery outburst of \bronnegen\ here.

\begin{table}
\caption{The new X-ray transient \bronnegen.}
\begin{threeparttable}
\begin{tabular*}{0.45\textwidth}{@{\extracolsep{\fill}}lc}
\hline
Parameter & Value \\
\hline
$N_{\mathrm{H}}$ ($\nh$) \dotfill & $(1.1\pm0.7)~\times10^{23}$  \\ 
$\Gamma$  \dotfill & $1.1 \pm 0.7$  \\ 
$F_{\mathrm{X}}$ ($\flux$)  \dotfill & $(6.6 \pm 2.0) \times 10^{-11}$ \\  
$L_{\mathrm{X}}$ ($\lum$)  \dotfill & $(5.1 \pm 1.5) \times 10^{34}$ \\
$L_{\mathrm{X}}^{\mathrm{peak}}$ ($\lum$) \dotfill & $1.1\times 10^{35}$  \\ 
$t_{\mathrm{ob}}$ (weeks) \dotfill &  1--2 \\ 
$f$ ($\fluence$) \dotfill &   $(3-9)\times10^{-6}$ \\ 
$\dot{M}_{\mathrm{ob}}$ ($\mdot$) \dotfill &   $1.3\times10^{-11}$ \\ 
$\langle \dot{M}_{\mathrm{long}} \rangle$ ($\mdot$) \dotfill &   $<3\times10^{-13}$ \\ 
\hline
\end{tabular*}
\label{tab:new}
\begin{tablenotes}
\item[] Note. -- $F_{\mathrm{X}}$ is the average unabsorbed flux and $L_{\mathrm{X}}$ is the corresponding luminosity for $D=8$~kpc (2--10 keV). The outburst fluence $f$ is the product of $F_{\mathrm{X}}$ and the outburst duration $t_{\mathrm{ob}}$. $\dot{M}_{\mathrm{ob}} $ is the estimated mass-accretion during outburst and $\langle \dot{M}_{\mathrm{long}}  \rangle$ the long-term averaged mass-accretion rate (for details, see \citep[][]{degenaar09_gc}). Errors are 90\% confidence. 
\vspace{-0.4cm}
\end{tablenotes}
\end{threeparttable}
\end{table}

\begin{figure*}[tb!]
 \begin{center}
\includegraphics[width=5.9cm]{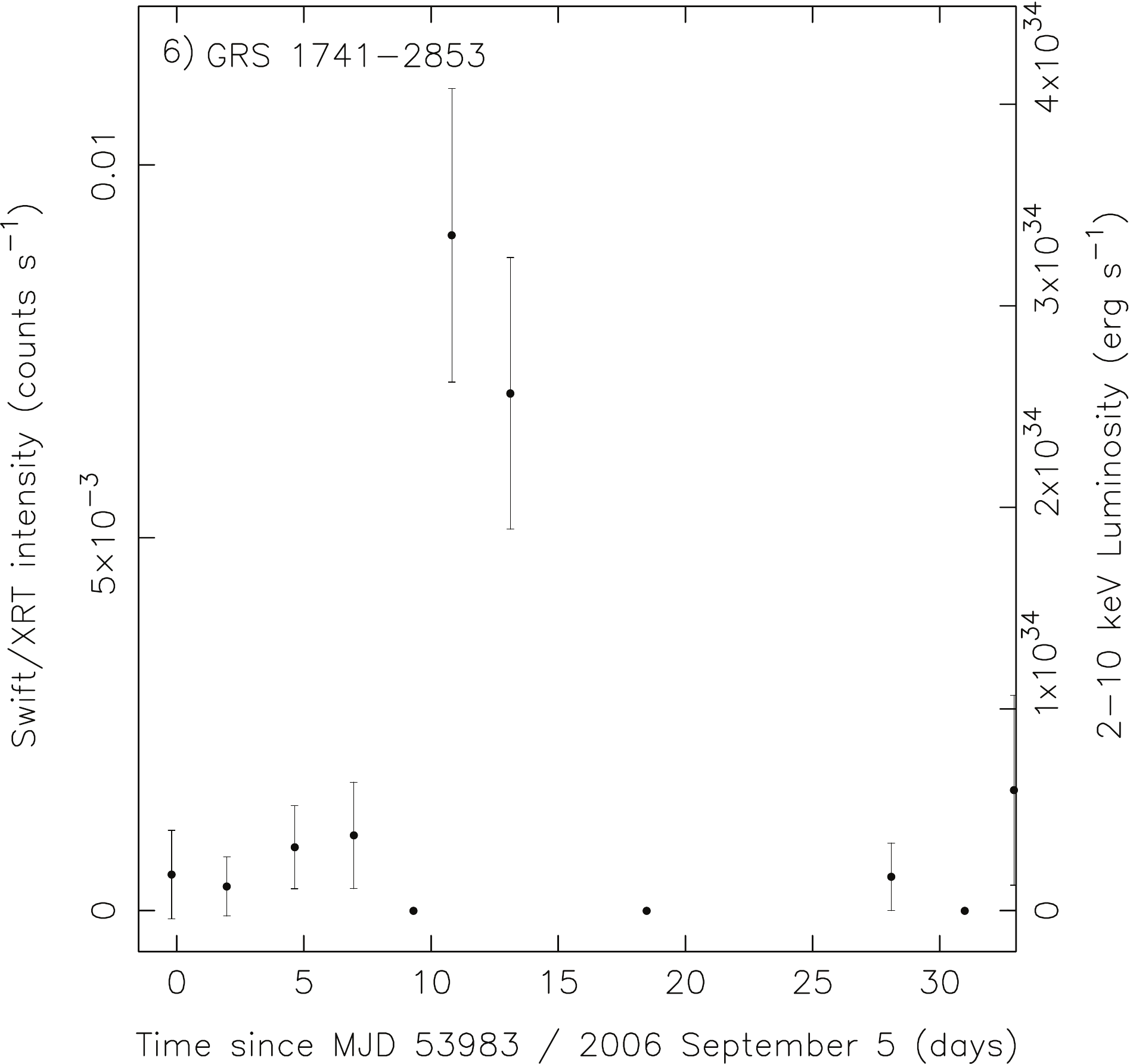}\hspace{+0.8cm}
\includegraphics[width=5.9cm]{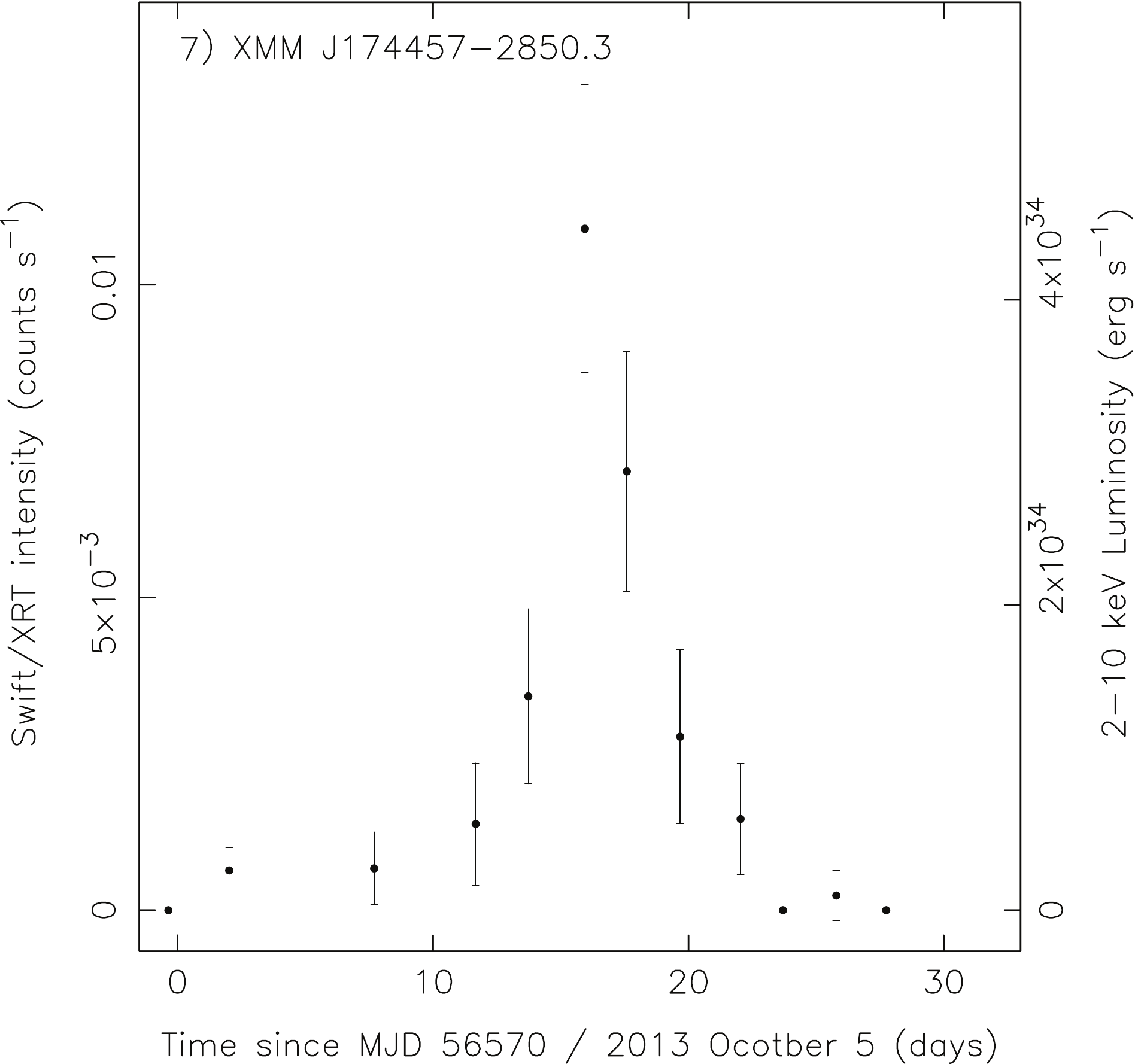}\vspace{+0.2cm}
\includegraphics[width=5.9cm]{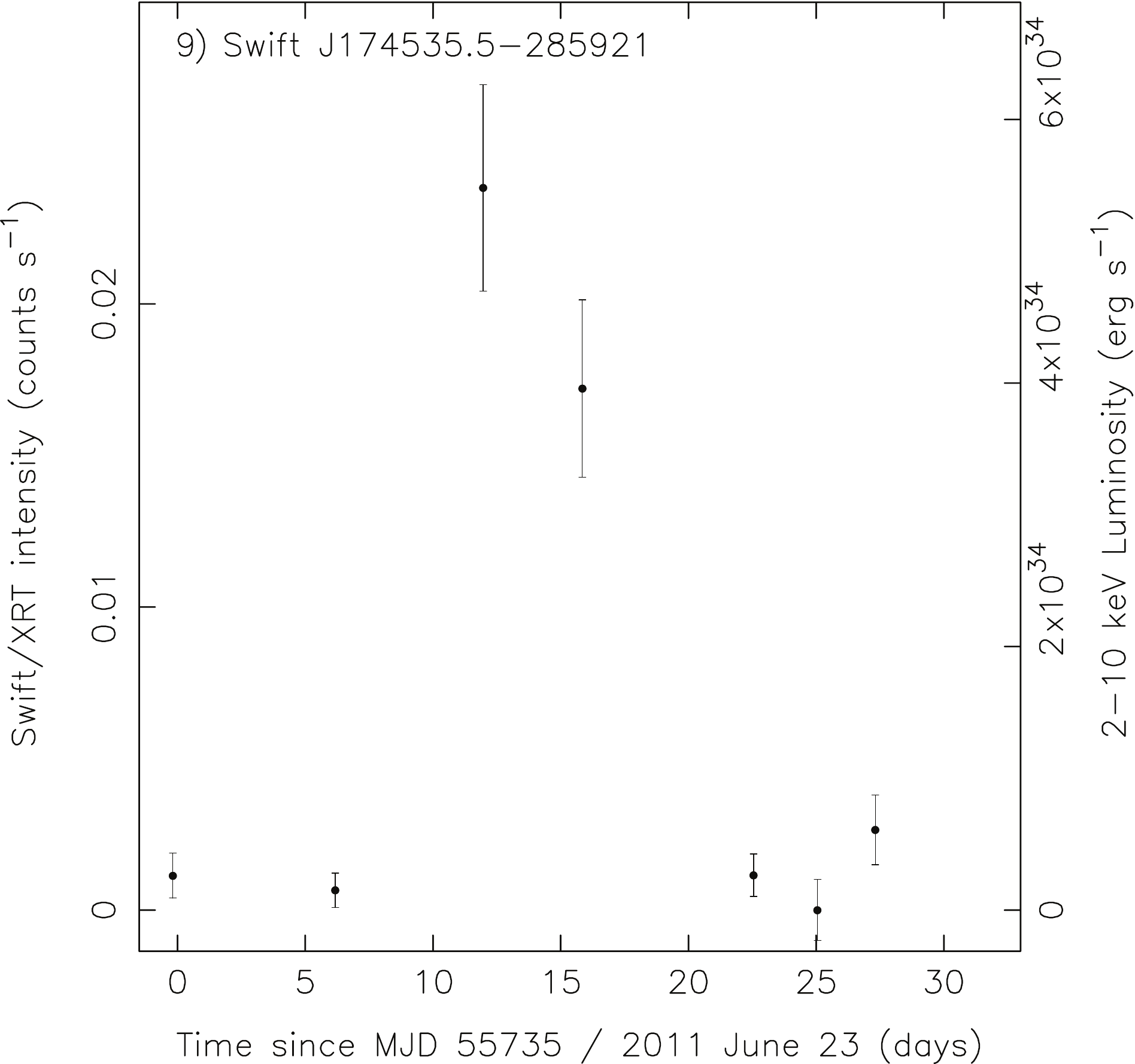}\hspace{+0.8cm}
\includegraphics[width=5.9cm]{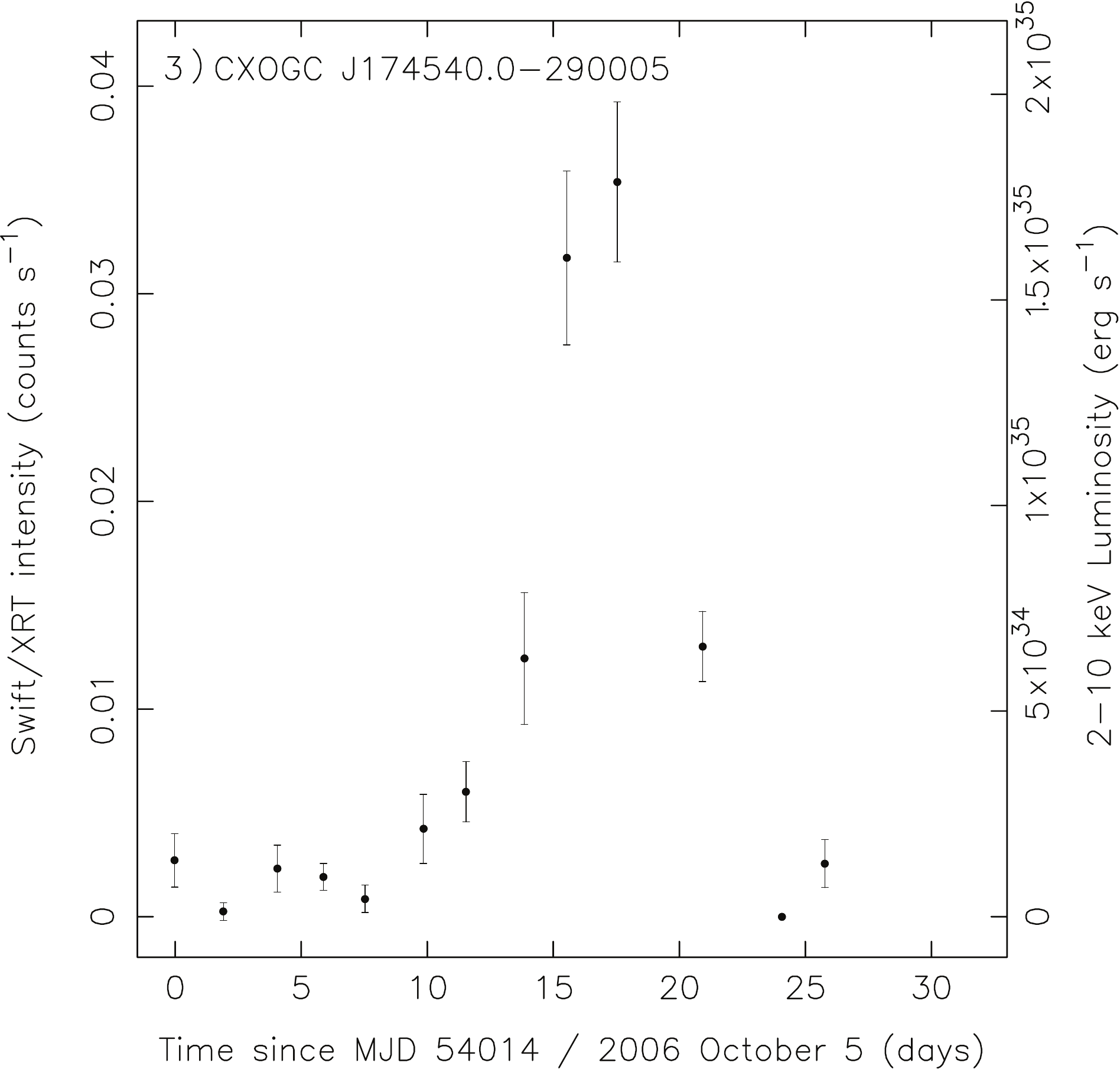}\vspace{-0.6cm}
    \end{center}
\caption[]{{Comparison between the light curves of low-level accretion activity in the bright neutron star LMXBs \grsbron\ (2006) and \xmmbron\ (2013) on top, and illustrative examples of the light curves of VFXBs in outburst, \bronnegen\ (2011) and \brondrie\ (2006), at the bottom. All displayed light curves were binned per two observations. Distances were assumed to be $D=8$~kpc for \brondrie\ and \bronnegen, but based on X-ray burst analysis we took $D=6.7$~kpc for \grsbron\ \citep[][]{trap2010} and $D=6.5$~kpc for \xmmbron\ \citep[][]{degenaar2014_xmmsource}.}}
\vspace{-0.5cm}
 \label{fig:lowlevel}
\end{figure*}

\vspace{-0.2cm}
\paragraph{\new} This transient X-ray source is located $\simeq1.3'$ NE of \sgra\ (see Figure~\ref{fig:image}) and was first detected on 2011 July 3, while it had not been seen prior to June 30 \citep[][]{degenaar2011_newtransient}. The source remained visible in subsequent observations performed on July 6, 8, and 9, was not  seen on July 15, 17, and 18, but reappeared during a single observation on July 21 \citep[][]{degenaar2011_newtransient2}. The source remained dormant after that (July 24 onwards). The long-term XRT light curve of this new X-ray transient is shown in Figure~\ref{fig:transientlc}, whereas a zoom of its 2011 outburst is shown at the bottom left in Figure~\ref{fig:lowlevel}. The results of fitting the average outburst spectrum (using obsID 91095027--29,  91095032, and 35650234; 5.1 ks of data), are listed in Table~\ref{tab:new}.

\chan\ observations performed on July 21 (i.e., during the re-flare) provided a sub-arcsecond localization of the new Galactic transient and allowed for the identification of the likely quiescent counterpart: \bronnegencxo\ \citep[][]{chakrabarty2011}. This is a faint object that was persistently detected at an intensity of $L_{\mathrm{X}}\simeq 2 \times 10^{31}~\lum$ during a \chan\ monitoring campaign of the Galactic center carried out in 2005--2008 (source number 1680; \citep[][]{muno2009}). 

The main outburst of \bronnegen\ had a duration of $6\lesssim t_{\mathrm{ob}}\lesssim15$~days, whereas the small re-flare cannot have been longer than 6 days. The average luminosity during the main outburst was $L_{\mathrm{X}}\simeq 5 \times 10^{34}~\lum$, whereas we estimate $L_{\mathrm{X}}\simeq 8 \times 10^{33}~\lum$ for the July 21 observation (see also \citep[][]{chakrabarty2011}). It seems likely that most of the energy was released during the main outburst. With an estimated outburst peak of $L_{\mathrm{X}}^{\mathrm{peak}}\simeq 1 \times 10^{35}~\lum$, \bronnegen\ falls into the class of VFXBs, just like the other two newly discovered transients \bronvier\ and \bronvijf\ (see Table~\ref{tab:spec}). 

From the observed outburst properties we can estimate the long-term average mass-accretion rate onto the compact primary (see Section~\ref{subsec:vf}). We estimate an average mass-accretion rate during outburst of $\dot{M}_{\mathrm{ob}} = RL_{\mathrm{acc}}/GM \simeq 1.3\times10^{-11}~\mdot$. In this relation $L_{\mathrm{acc}}\simeq3 L_{\mathrm{X}}$ is the estimated bolometric accretion luminosity \citep[][]{zand07}, $G$ is the gravitational constant, and $R$ and $M$ are the radius and mass of the compact primary, respectively. Here we assumed a neutron star primary with $R=10$~km and $M=1.4~\Msun$ (see \citep[][]{degenaar09_gc} for caveats). 

The long-term average accretion rate can be estimated as $\langle \dot{M}_{\mathrm{long}}  \rangle $$= $$\dot{M}_{\mathrm{ob}} \times t_{\mathrm{ob}}/t_{\mathrm{rec}}$, where $t_{\mathrm{rec}}$$=$$t_{\mathrm{ob}}+t_{\mathrm{q}}$ is the recurrence time, i.e., the sum of the quiescent and outburst intervals. In 2014 the source position was covered for 39 consecutive weeks with no data gaps exceeding 5 days. We can therefore take this as a lower limit on the time it spends in quiescence, i.e.,  $t_{\mathrm{q}}$$>$$39$~weeks and $t_{\mathrm{rec}}$$>$$40$~weeks. The duty cycle is then $t_{\mathrm{ob}}/t_{\mathrm{rec}}$$<$$0.022$ (2.2\%), giving $\langle \dot{M}_{\mathrm{long}}  \rangle$$<$$3\times10^{-13}~\mdot$. This low value may indicate that \bronnegen\ also harbors an old degenerate companion and a small orbit \cite[][]{degenaar09_gc,heinke2014_gc}.

\subsection{Low-level accretion in bright X-ray binaries}\label{subsec:low}
Remarkably, the \swift\ Galactic center monitoring campaign has also exposed very-faint X-ray activity from LMXBs that normally exhibit brighter outbursts. \ascabron, for instance, exhibited four different outbursts between 2006--2014; in 2007--2008 and 2013--2014 it showed long ($>$1.5~yr) and bright ($L_{\mathrm{X}}^{\mathrm{peak}}\simeq6-7\times10^{36}~\lum$) outbursts (Table~\ref{tab:spec}). However, its 2006 and 2010 outbursts were both shorter (few months) and an order of magnitude fainter (\citep[][]{degenaar09_gc,degenaar2014_gctransients}, see also Figure~\ref{fig:transientlc}). Moreover, in 2006 \swift\ uncovered a very brief ($\simeq$1~week) and very faint ($L_{\mathrm{X}}^{\mathrm{peak}}\lesssim7\times10^{34}~\distgrs~\lum$) outburst from \grsbron\ (see Figure~\ref{fig:lowlevel}), whereas it normally displays weeks-long outbursts that peak at $L_{\mathrm{X}}^{\mathrm{peak}}\simeq\times10^{36-37}~\distgrs~\lum$ (\citep[][]{degenaar09_gc}, see Table~\ref{tab:spec}). 

A strikingly similar mini-outburst as that from \grsbron\ was detected with \swift\ from \xmmbron\ in 2013; a clear excess of photons was found at the source position in five consecutive observations performed on October 19, 20, 21, 22, and 23, whereas it went undetected on October 18 and 24, i.e., $t_{\mathrm{ob}}$$=$5--6~days. The averaged spectrum of these five observations can be described by an absorbed power-law model with an index of $\Gamma$$=$$0.8\pm0.8$ (for a fixed $N_{\mathrm{H}}$$=$$1.1 \times 10^{23}~\nh$; \citep[][]{degenaar2014_xmmsource}), yielding a 2--10 keV unabsorbed flux of $F_{\mathrm{X}}$$ = $$(5.9\pm2.2)\times 10^{-12}~\flux$. This translates into an average luminosity of $L_{\mathrm{X}} $$=$$(3.0\pm1.1)\times 10^{34}~\distxmm~\lum$, whereas we estimate a peak luminosity of $L_{\mathrm{X}}^{\mathrm{peak}} $$=$$ 4.5\times 10^{34}~\distxmm~\lum$. The light curve of this short, faint outburst of \xmmbron\ is shown in Figure~\ref{fig:lowlevel} (top right), where it is compared to mini-outburst of \grsbron\ in 2006 (top left; \citep[][]{degenaar09_gc}).

For comparison, we also show in Figure~\ref{fig:lowlevel} two example light curves of the activity of VFXBs; the 2006 outburst of \brondrie\ (lower right), and the 2011 discovery outburst of \bronnegen\ (lower left). This plot illustrates that the mini-outbursts of the otherwise bright LMXBs are very similar in terms of duration, peak intensity, and energetics to some outbursts of VFXBs. Since \grsbron\ and \xmmbron\ also show longer and brighter outbursts, it is likely that their mini-outbursts are the result of accreting only a (small) portion of the entire  disk \citep[][]{degenaar2010_gc,heinke2014_gc}. Similar events have also been detected from a few other bright neutron star LMXBs (e.g., \xte, \ksbron, \saxrudy, and \maxisource; \citep[][]{fridriksson2011,degenaar2013_ks1741,wijnands2013,homan2014}). It can therefore be hypothesized that some of the VFXBs could be bright LMXBs that exhibited low-level accretion activity (e.g., \citep[][]{wijnands2012,heinke2014_gc}, see also \citep[][]{fridriksson2011}).

\subsection{LMXB/millisecond radio pulsar transitional object}\label{subsec:trans}
Apart from short, faint accretion outbursts such as seen in 2013 (see Figure~\ref{fig:transientlc} and Section~\ref{subsec:low}), the \swift\ Galactic center monitoring program revealed that \xmmbron\ also displays much longer (months) periods of sustained low-level activity, sometimes without an associated outburst \citep[][]{degenaar2014_xmmsource}. In fact, combining the \swift\ data with archival \chan\ and \xmm\ observations, it was demonstrated that the source appears to exhibit three different luminosity regimes; 1) it regularly displays accretion outbursts with $L_{\mathrm{X}} \simeq10^{34}-10^{36}~\distxmm~\lum$, 2) on rare occasions it has been found in deep quiescence with $L_{\mathrm{X}}\lesssim10^{33}~\distxmm~\lum$, and 3) most often it is detected at an intermediate regime of $L_{\mathrm{X}}\simeq10^{33}-10^{34}~\distxmm~\lum$. There appears to be little variation in the X-ray spectrum despite this orders of magnitude change in luminosity \citep[][]{degenaar2014_xmmsource}. 

This behavior is difficult to explain as thermal-viscous instabilities in the accretion disk, since the mass-accretion rate during these low-activity periods should not be sufficient to irradiate the disk and keep the mass-flow going \citep[][]{degenaar2010_gc}. However, the detection of an energetic Type-I X-ray burst in 2012 makes it unlikely that the neutron star is feeding off the wind of its companion, hence increasing the mystery of its peculiar X-ray properties \citep[][]{degenaar2014_xmmsource}.  

The flux variability of \xmmbron, and the lack of strong spectral changes between different luminosity states, is remarkably similar to that of a neutron star in the globular cluster M28, \psrm\ \citep[][]{linares2014}. This source was known to be a millisecond radio pulsar (MSRP), but in 2012 suddenly exhibited an accretion outburst like that seen LMXBs \citep[][]{papitto2013_nature}. It had long been suspected that MSRPs are the descendants of LMXBs (e.g., \citep[][]{alpar1982,bhattacharya1991,strohmayer1996,wijnands1998}), and the discovery of a neutron star directly switching between these two different manifestations has opened up a new opportunity to investigate their evolutionary link. Two other MSRPs display similar behavior, \psr\ (e.g., \citep[][]{archibald2009,archibald2014,stappers2013,patruno2014,deller2014}) and \xss\ (e.g., \citep[][]{demartino2013,bassa2014,roy2014,papitto2014_puls}), and together these three are referred to as the `MSRP/LMXB transitional objects'. Discovering more of such neutron stars, either drawn from the MSRP or from the LMXB population, is highly desired to further investigate their connection. Based on seven years of \swift\ Galactic center monitoring, we demonstrated that the peculiar neutron star \xmmbron\ may be another member of this class \citep[][]{degenaar2014_xmmsource}.

\section{Summary and outlook}\label{subsec:conclude}
We have reviewed the main discoveries of nine years of \swift\ X-ray monitoring of the center of the Milky Way Galaxy, in which short 1-ks exposures have been taken almost daily since 2006.  The \swift\ detection of seven bright ($L_{\mathrm{X}}$$\gtrsim$$10^{35}~\lum$) X-ray flares from \sgra\ constitutes more than half of the current number of such bright flares observed, and has allowed for an estimate of their recurrence time (once per 5--10 days) as well as a comparison of their spectral properties (hinting that flares of similar brightness may have different spectral shapes; Section~\ref{subsec:flares}). Moreover, the discovery of the magnetar \magnetar, which seems to be orbiting the supermassive black hole, has provided a measurement of the magnetic field near \sgra\ and has led to new observing strategies for finding other radio pulsars in the same region (Section~\ref{subsec:magnetar}). 

In addition, the unique combination of sensitive observations and dense sampling of the \swift\ program allowed for a first detailed, systematic study of the peculiar class of VFXBs, X-ray binaries with very dim outbursts and unusually low mass-accretion rates, providing more insight into their nature (Section~\ref{subsec:vf}). In nine years time, \swift\ discovered three new transient VFXBs near the Galactic center (Section~\ref{subsec:new}). Furthermore, \swift\ exposed that bright LMXBs can display low-level accretion activity that looks similar to the outbursts of VFXBs (Section~\ref{subsec:low}). Finally, owing to this \swift\ program it was found that \xmmbron\ displays peculiar X-ray flux variability and could be a member of the recently discovered class of LMXB/MSRP transitional objects, which hold great potential to further our knowledge of the evolutionary link between these two different neutron star manifestations (Section~\ref{subsec:trans}). 

Building and following-up on this exciting series of discoveries, continuing this \swift\ legacy allows us to:

\begin{enumerate}
\item \textbf{Continue catching \sgra\ X-ray flares}\\
Confirming that X-ray flares from \sgra\ can have different spectral shapes would give important insight into their underlying emission and production mechanism (e.g., \citep[][]{degenaar2013_sgra,barriere2014}). \swift\ has provided a first glimpse in this direction, by finding that one of the six flares detected in 2006--2011 may have had a different spectral shape \citep[][]{degenaar2013_sgra}. However, the result was of only low significance, so further confirmation is required. By continuing the \swift\ monitoring campaign we can collect more X-ray flares from \sgra\ and increase the sample available for a comparative study.  

\item \textbf{Keep watch for changing activity of Sgr A$^*$}\\
Now that G2 has swung by \sgra\ (\citep[][]{witzel2014,pfuhl2015}), its possible disruption and accretion could increase the X-ray emission of the supermassive black hole. Dedicated monitoring and target-of-opportunity programs remain in line to study this rare phenomenon. It is widely recognized that \swift's Galactic center monitoring program is of vital importance to promptly detect any changes in the (persistent or flaring) emission from \sgra, and to trigger follow-up observations with other observatories and at other wavelengths. 

\item \textbf{Further study very-faint accretion outbursts}\\\
In 2006--2014, \swift\ detected 16 very-faint accretion outbursts from 9 X-ray binaries. The daily, sensitive XRT observations have yielded a wealth of information about their outburst profiles and energetics, which cannot be obtained with any other X-ray satellite. Continuation of this program is guaranteed to capture new outbursts of VFXBs, giving further insight into the nature of this peculiar (yet perhaps dominant) sub-class of X-ray binaries.

\item \textbf{Test the nature of \xmmbron}\\
A distinguishing property of the recently discovered LMXB/MSRP transitional objects is that these occupy different X-ray luminosity regimes, most likely governed by the interaction between the accretion flow and the magnetic field (e.g., \citep[][]{linares2013_M28,patruno2014,bassa2014}). By virtue of the \swift\ Galactic center monitoring campaign, we found that \xmmbron\ may display similar behavior \citep[][]{degenaar2014_xmmsource}. However, due to its location $\simeq$$14'$ from \sgra, it was only covered during a fraction of the time. In 2015--2016 the observations will be offset by $\simeq2'$ from the nominal aimpoint, to ensure that \xmmbron\ is covered by every pointing. Following this source daily for a year allows us to establish if it indeed exhibits three distinct accretion regimes, strengthening its LMXB/MSRP classification.

\end{enumerate}

\vspace{-0.4cm}
\section*{Acknowledgements}
ND acknowledges support via an EU Marie Curie Intra-European fellowship no. FP-PEOPLE-2013-IEF-627148.

\end{document}